\let\LaTeXcline\cline
\documentclass[pdflatex,sn-mathphys-num,iicol]{sn-jnl}
\let\cline\LaTeXcline

\usepackage{amsmath}
\usepackage{amssymb}
\usepackage{bm}
\usepackage{textcomp}
\usepackage[english]{babel}
\usepackage[utf8]{inputenc}
\usepackage{times}
\usepackage{cancel}
\usepackage[T1]{fontenc}
\usepackage{amsfonts}
\usepackage{soul}
\usepackage{gensymb}
\usepackage{multirow}
\usepackage[normalem]{ulem}
\usepackage{xcolor}
\usepackage{cancel}

\usepackage{textcomp}
\usepackage{graphicx, subfig}
\usepackage[justification=raggedright]{caption}

\DeclareCaptionLabelFormat{name}{}

\usepackage{tikz}
\usepackage{array}
\usepackage{xcolor}
\definecolor{lightgray}{gray}{0.8}
\usepackage{ dsfont }
\usepackage{ bbold }

\renewcommand{\vec}[1]{\bm{#1}}
\renewcommand{\hat}[1]{\bm{\widehat{#1}}}

\newcommand{\update}[1]{\textcolor{black}{#1}}
\newcommand{\UpdatE}[1]{\textcolor{black}{#1}}

\newcommand{\bn}{\textbf{n}}
\newcommand{\br}{\textbf{r}}
\newcommand{\bu}{\textbf{u}}
\newcommand{\eref}[1]{(\ref{#1})}
\newcommand{\Fref}[1]{Fig.~\ref{#1}}

\begin{document}
\title{Quorum sensing of light-activated colloids

in nematic liquid crystals \vspace{0.5cm}}

\author*[1]{\sur{Antonio Tavera-V\'{a}zquez}}\email{atv.tavera@gmail.com }

\author[2,3]{\sur{David Martin}}

\author[4]{\sur{Haijie Ren}}

\author[1]{\sur{Sam Rubin}}

\author[1,5]{\sur{Andr\'{e}s~C\'{o}rdoba}}

\author[4]{\sur{Rui Zhang}}

\author[2,6]{\sur{Vincenzo Vitelli}}

\author*[1,7,8,9]{\sur{Juan J. de Pablo}}\email{jjd8110@nyu.edu }

\affil*[1]{\orgdiv{Pritzker School of Molecular Engineering}, \orgname{University of Chicago},
\orgaddress{%\street{South Ellis Avenue},
\city{Chicago}, \postcode{60637}, \state{IL}, \country{USA}}}

\affil[2]{\orgdiv{Leinweber Institute for Theoretical Physics}, \orgname{University of Chicago}, 
\orgaddress{%\street{South Ellis Avenue},%
\city{Chicago}, \postcode{60637}, \state{IL}, \country{USA}}}

\affil[3]{\orgdiv{LPTMC}, \orgname{Sorbonne Universit\'e \& CNRS},
\orgaddress{%\street{South Ellis Avenue},
\city{Paris}, \postcode{75252}, \country{France}}}

\affil[4]{\orgdiv{Department of Physics}, \orgname{Hong Kong University of Science and Technology}, 
\orgaddress{%\street{Kowloon},%
\city{Hong Kong}, \postcode{10587}, \state{Hong Kong}, \country{China}}}

\affil[5]{\orgdiv{\'{A}rea de Ciencias Fundamentales, 
Escuela de Ciencias Aplicadas e Ingenier\'{i}a}, \orgname{Universidad EAFIT}, 
\city{Medell\'{i}n}, \postcode{050022}, \country{Colombia}}

\affil[6]{\orgdiv{James Franck Institute}, \orgname{University of Chicago},
\city{Chicago}, \postcode{60637}, \state{IL}, \country{USA}}

\affil*[7]{\orgdiv{Chemical and Biomolecular Engineering, Tandon School of Engineering}, \orgname{New York University},
\orgaddress{%\street{South Ellis Avenue},
\city{Brooklyn}, \postcode{11201}, \state{NY}, \country{USA}}} 

\affil*[8]{\orgdiv{Courant Institute of Mathematical Sciences}, \orgname{New York University},
\orgaddress{%\street{South Ellis Avenue},
\city{Manhattan}, \postcode{10012}, \state{NY}, \country{USA}}}

\affil*[9]{\orgdiv{Department of Physics}, \orgname{New York University},
\orgaddress{%\street{South Ellis Avenue},
\city{Brooklyn}, \postcode{11201}, \state{NY}, \country{USA}}}

%\affil[5]{\orgdiv{Materials Science Division}, \orgname{Argonne National Laboratory}, \orgaddress{\city{Lemont}, \postcode{60439}, \state{Illinois}, \country{USA}}}

\abstract{
%%%%%%%% version 1 about quorum-sensing %%%%%%%%
Motile living organisms routinely probe their surroundings to adapt in ever-evolving environments. 
Although synthetic microswimmers offer surrogates for self-propelled living entities, they often lack the complex feedback mechanisms that enable organisms to adapt.  
In this work, we present an experimental platform in which light-activated colloids dispersed in a nematic liquid crystal can (i) switch from directed to active Brownian motion depending on the nematic anchoring and (ii) mechanically adjust their motility in response to crowding, effectively enforcing quorum-sensing interactions.
Both features are caused by a distinctive self-propulsion mechanism as unveiled through experiments, simulations, and theory. 
%which stems from immersing the colloids in a nematic liquid crystal.
% : the inner drive controls the mesogens' configuration around the colloid.
We characterize the dynamics of a single colloid and demonstrate that its motion is captured by an active Brownian particle model if the nematic anchoring is homeotropic, and by directed self-propulsion along the nematic director if the anchoring is planar.
Next, we investigate the many-body dynamics, showing that it undergoes a clustering phase separation through effective quorum-sensing interactions.
%Our results are supported by numerical simulations of both the surrounding liquid crystal and our active-matter dynamical model. 
Our work suggests how to create adaptive materials with life-like capabilities using readily accessible properties of liquid crystals and colloids without explicitly engineering any of the needed mechano-chemical feedbacks.

}


\maketitle
The survival of living organisms depends on their ability to sense their surroundings and react adaptively: bacteria perform chemotaxis to forage food sources \cite{adler1975chemotaxis,berg1975chemotaxis}, plant growth is regulated by neighboring foliage density \cite{hautier2009competition}, and birds adapt their flight to remain in the flock \cite{ballerini2008interaction}.
A minimal model for such self-regulation is quorum sensing, where entities modify their dynamics in response to interactions with their neighbors \cite{miller2001quorum,cates2015motility}.
Although synthetic active particles have proved a controlled testbed for studying the collective behavior of living entities, they are typically made of inanimate beads that lack the ability to perform quorum-sensing interactions \cite{howse2007self, theurkauff2012dynamic,palacci2013living,bricard2013emergence,zottl2016emergent,zhang2017active}. 
Until now, the study of self-regulation mechanisms and their corresponding many-body emergent behavior has focused on computer-controlled active units \cite{lavergne2019group,fernandez2020feedback,muinos2021reinforcement,ben2023morphological} or dense assemblies of Quincke rollers \cite{geyer2019freezing} and rods \cite{lefranc2025quorum}.

% so minimal that they remain, at the exception of a few examples, 
In this paper, we report how Janus particles evolving in a nematic liquid crystal (LC) spontaneously self-propel and perform quorum-sensing by light activation.
We show how the structured anisotropic LC enables control over the dynamics of the colloids, directing them in a specific direction and endowing them with additional activity regulation through quorum-sensing.
Our efforts are fourfold.
First, we describe the experimental setup, detailing the interplay between the mesogens and the Janus particles, be it in the presence or in the absence of light. We further validate our experimental observations with continuum numerical simulations of the LC.
Second, we characterize the non-equilibrium mechanism behind the emergence of self-propulsion through a qualitative derivation further supported by numerical simulations.  
Third, we probe the dynamics of a single Janus colloid and propose corresponding phenomenological models inspired by active matter.
Lastly, we explore the many-body dynamics, showing that colloids perform an effective quorum-sensing slow down, which ultimately leads to the emergence of a clustered phase. 

% The particle trajectories were tracked at different intensities of the applied light. 
% The model suggests the possibility of utilizing Janus particles as probes for measuring the rheology of nematic LCs. 
% Our simple colloidal system provides an experimental platform for studying the impact of activity on transport mechanisms in anisotropic environments, thereby paving the way for the conception of synthetic surrogate to complex active biomaterials. 
% This paving the way for the conception of an experimental platform for highly structured anisotropic active materials.

% Our experimental design relies on altering the Janus particle Saturn ring when inducing a local nematic-isotropic phase transition. 

\begin{figure*}[h]
\centering
\includegraphics[width=\textwidth]{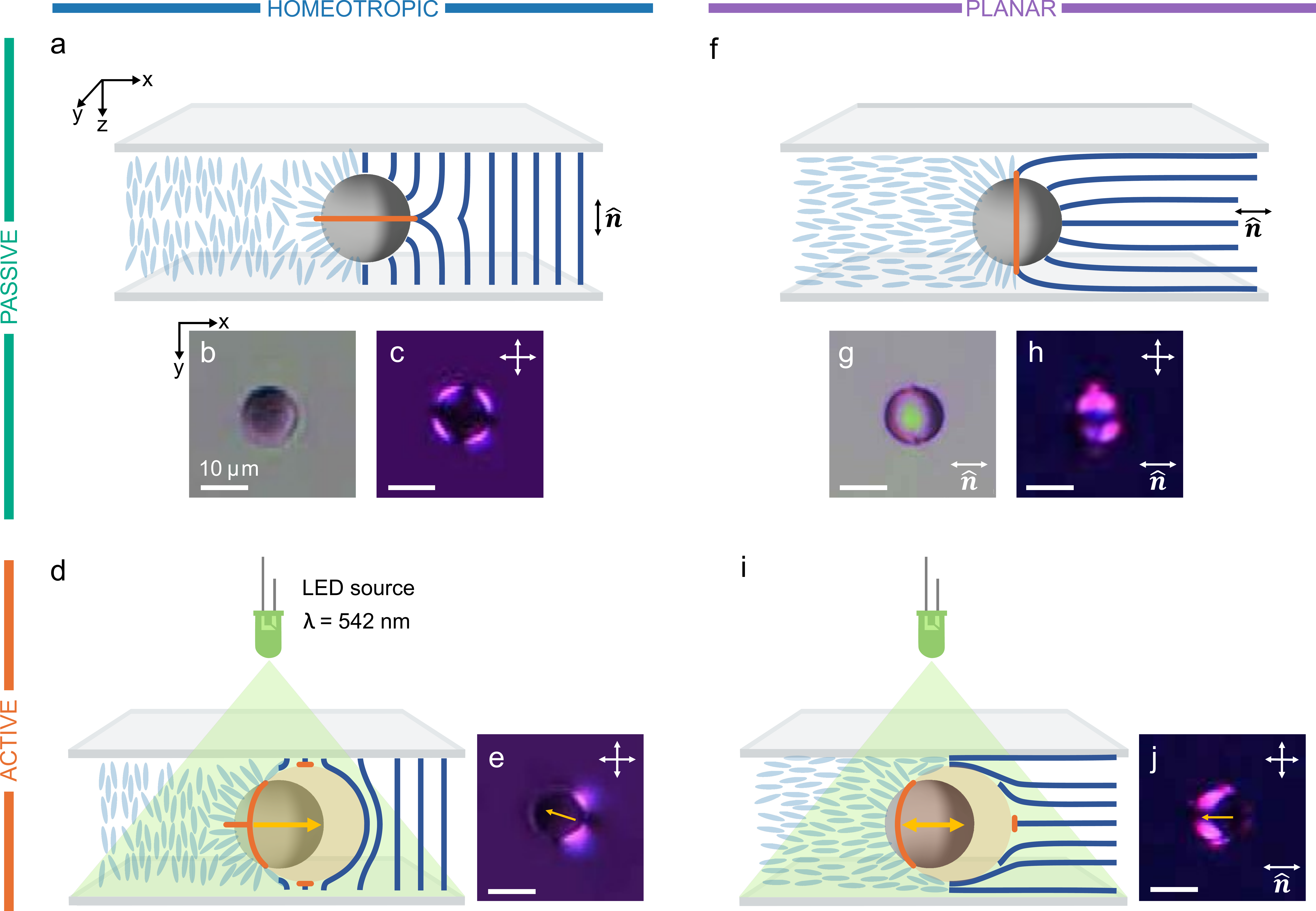}
\vspace{0.2cm}
\caption{
Visualization of the LC configuration around a Janus colloid with homeotropic (left) and planar (right) anchoring at the confining cell, as well as in the absence (top) or presence (bottom) of activation light. \textbf{a,f,d,i} Schematic view of the mesogens around the colloid. The darker area (right side) marks the coated side of the particle while the region thinly colored in orange indicates the bubble of the isotropic phase. 
The unit vector $\hat{\bm{n}}$ is the bulk's nematic director, blue lines represent the nematic field, while orange lines indicate topological defects.
Finally, the yellow arrow indicates the potential directions of the emerging motility: in the planar case, the colloid can self-propel toward or away from the coated side.
% Note that the mesogens attach homeotropically to both the particle's and the bubble's surfaces. 
\textbf{b,g} Brightfield experimental snapshots of a Janus particle. \textbf{c,e,h,j} Cross-polarized experimental snapshots of a Janus particle. 
In \textbf{c,} the birefringent bright lobes are characteristic of a Saturn ring located in the $xy$ plane. 
% In \textbf{h,} the rear lobes are preserved.
In \textbf{h,j,} small bumps at the particles' surface suggest the presence of Saturn rings in the $yz$ plane. 
% Finally, in \textbf{e,j,} only two out of the four lobes are preserved, which indicates the presence of 
All scale bars represent 10 $\mu$m.
}

\label{fig:intro_mechanism}
\end{figure*}

\section*{Experimental setup}\label{experimental}

Our platform consists of a rectangular cell with dimensions $2\times2$ cm and height of 12 $\mu$m filled with a suspension of silica spheres $10$ $\mu$ m, half coated with a light absorbent titanium layer.
A single LED source that shines from above throughout the cell controls the light intensity (see \Fref{fig:intro_mechanism}).
Upon illuminating a Janus colloid, its metal layer heats up the surrounding LC and induces a local nematic to isotropic transition, resulting in the formation of a disordered LC bubble attached to the coated side.
Such a configuration frustrates the orientation of the mesogens at the contact line between the bubble and the colloid because the anchoring is homeotropic on silica but approximately planar on the bubble's surface. 
Hence, nematic defects emerge in order to ensure the continuity of the director field near the contact line. 
As detailed in the next section, the presence of these defects leads to a net self-propulsion force which, depending on the experimental conditions, we found to be either parallel or antiparallel to the bead's revolution axis (see \Fref{fig:intro_mechanism}d and i). 

We studied two different treatments of the cell surface resulting in homeotropic and planar anchoring of the mesogens (details for their experimental manufacturing can be found in the Methods). As shown in \Fref{fig:intro_mechanism}, these two distinct orientations at the boundary deeply impact the configuration of the defects around the colloid, since the continuity of the director field must be preserved everywhere. 
Consequently, the defect structures are affected by the anchoring and so is the resulting dynamics of the colloid.
In the remainder of this paper, we will thus distinguish between these two scenarios, namely homeotropic and planar anchoring, and we will specify our results for each of these two cases.
% We now report our experimental characterization of the LC nematic field around a single Janus colloid. 

% \subsection{Homeotropic anchoring}

Homeotropic confinement, as shown on the left side of \Fref{fig:intro_mechanism}, induces a nematic order along the $z$ direction far from the colloid (see \Fref{fig:intro_mechanism}a). 
Close to the bead, the LC's configuration is light-dependent: the passive, non-illuminated case is displayed in the upper row of \Fref{fig:intro_mechanism} while the active, illuminated case is shown in the lower row. 
\\
\textit{Passive:} Cross-polarized microscopy (\Fref{fig:intro_mechanism}c) exhibits a birefringence pattern with four bright lobes, thus confirming the homeotropic anchoring of the mesogens on the surface of the bead.
The frustration between the bulk's mesogens and the director field close to the colloid generates a Saturn ring defect located in the bead's equatorial plane, parallel to the cell's surfaces (see orange line in \Fref{fig:intro_mechanism}a) \cite{gu2000observation}.
\\
\textit{Active:} The light-generated bubble of isotropic LC breaks the rotational symmetry of the nematic field around $z$ (see the \Fref{fig:intro_mechanism}d sketch).
As shown in \Fref{fig:intro_mechanism}e, cross-polarized microscopy only reports two birefringence lobes located on the bare or non-coated hemisphere, confirming an anchoring change of the mesogens on the bubble's surface.
This generates two additional regions of geometric frustration for the director field that were absent in the passive case.
The first one is located at the contact line between the bubble and the colloid, where planarly-anchored mesogens must encounter their homeotropically-anchored counterpart. 
This first source of frustration is relaxed by the formation of an off-centered Saturn ring at the contact line ($yz$ plane), as shown in \Fref{fig:intro_mechanism}d.
The second region of frustration is located at the top and bottom tips of the bubble, where mesogens strongly favor homeotropic anchoring at the cell's surface. 
This second source of frustration is relaxed by the formation of two boojum defects \cite{mermin1981boojums,tasinkevych2012liquid}, as shown in \Fref{fig:intro_mechanism}d.
% The additional frustration between this planar anchoring and the homeotropic mesogens on the non-coated hemisphere generates a Saturn ring of defects at the contact line between the bubble and the colloid, as shown in \Fref{fig:intro_mechanism}d.
Note that the colloid always self-propels toward the coated hemisphere.
% \\
% \textbf{Planar anchoring}
% \\
% \subsection{Planar anchoring}

Planar anchoring, as shown on the right hand side of \Fref{fig:intro_mechanism}, induces a nematic order along the $x$ direction in the bulk (see \Fref{fig:intro_mechanism}f).  
Similarly to homeotropic confinement, the behavior of the mesogens close to the bead is light dependent.
\\
\textit{Passive:} In brightfield microscopy (see (\Fref{fig:intro_mechanism}g)), we observe two lumps located at the top and bottom of the colloid. These lumps are characteristic of a Saturn ring located in the $yz$ plane (orange line in \Fref{fig:intro_mechanism}f) whose presence is further confirmed in cross-polarized microscopy (\Fref{fig:intro_mechanism}h).
This defect ring stems from the frustration between the homeotropically anchored mesogens at the surface of the bead and their planar anchored counterparts at the cell's surface.
\\
\textit{Active:} 
% \Fref{fig:intro_mechanism}i sketches the LC's configuration around the activated particle. 
The light-generated isotropic bubble forces the reconfiguration of the mesogens (see \Fref{fig:intro_mechanism}i). 
Cross-polarized microscopy (\Fref{fig:intro_mechanism}j) exhibits the two characteristic lumps of a Saturn ring located at the contact line between the isotropic bubble and the colloid (see the orange line in \Fref{fig:intro_mechanism}i).
A new source of frustration also emerges at the right tip of the bubble, since the bulk mesogens must continuously distort to match the planar anchoring at the nematic-isotropic (NI) interface.
This second source of frustration is relaxed by the emergence of a Bojuum defect located at the right tip of the bubble, as shown in \Fref{fig:intro_mechanism}i.
Note that the particle's motility always emerges parallel to the bulk nematic director ($x$ axis), but can be directed toward or away from the coated hemisphere, depending on the light intensity.

\begin{figure*}
\centering
\includegraphics[width=0.82\linewidth]{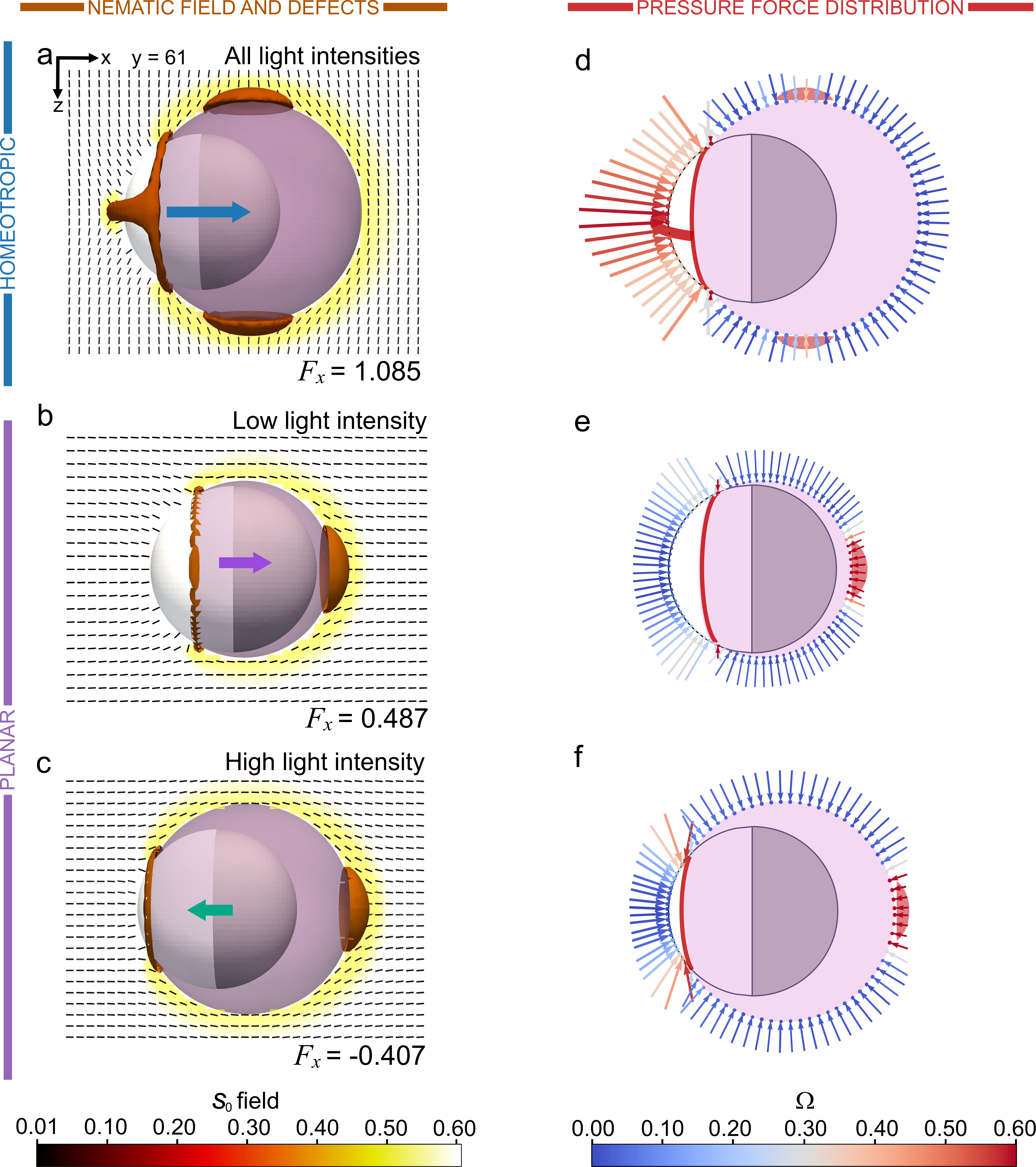}
\vspace{0.5cm}
\caption{Continuum simulations of the colloid-isotropic bubble system for homeotropic (top) and planar (bottom) anchoring, showing the nematic field with its topological defects (left) as well as the pressure force distribution (right). 
\textbf{a,b,c} Topological defects are indicated by orange surfaces, the isotropic bubble is represented by the lightly opaque magenta color, located under the darker side of the colloid, which indicates its coated side. 
The colormap shows the magnitude of the scalar order parameter field $S_0$, and the arrow indicates the direction of the net self-propulsion force $F_x$.
In \textbf{a,} an off-centered Saturn ring is located at the contact line between the isotropic bubble and the colloid, as an extension of the Saturn ring that remains at the $xy$ plane of the non-coated hemisphere. Two other localized topological point-like defects are further present at the bottom and at the top of the isotropic bubble.
In \textbf{b,c,} an off-centered Saturn ring is located at the contact line between the isotropic bubble and the colloid. Another localized topological point-like defect is situated at the right tip of the isotropic bubble. As is noticed, increasing the light intensity leads to an increased covering of the colloid by the isotropic bubble.
\textbf{d, e, f} Force distribution on the particle-bubble compound corresponding to panels \textbf{a}, \textbf{b}, and \textbf{c}. The colormap of the arrows, denoted as $\Omega$, indicates the magnitude of the force's component due to the presence of defects, namely $\bm{\sigma}^{\textrm{defect}}d\vec{S}$. It is defined as $\Omega = \big|\frac{\vec{F}_{sim}\cdot\vec{P}_s}{|\vec{F}_{sim}|^2}-1\big|$, with $\vec{F}_{sim}$ the numerically computed force while the pressure $P_s$ can take two values: $P_p$ on the particle surface or $P_b$ on the bubble's surface. Thus, $\Omega$ deviates from zero only when the local defects contribute to the force. Further details on simulations are provided in the Methods.}
\label{fig:simulations}
\end{figure*}

To support these experimental observations, we further performed numerical simulations of the LC around the colloid-bubble compound.
% \section{Continuum simulations}
As shown in \Fref{fig:simulations}, our simulations reveal the nematic director field $S_0$ with its defect structure (\Fref{fig:simulations}a-c) as well as the force distribution (\Fref{fig:simulations}d-f) at the surface of the system. 
The details of our non-equilibrium numerical approach, which uses two different temperatures in the distinct LC's phases, can be found in the Methods.
For both homeotropic and planar confinements, our simulations confirm the experimental configuration of the LC reported in \Fref{fig:intro_mechanism} and predict the direction of the emerging self-propulsion force.

% \subsection{Homeotropic anchoring}
For homeotropic confinement, Fig.\ref{fig:simulations}a shows the emergence of a vertical defect ring at the contact line between the particle and the isotropic bubble, following the continued ring in the horizontal equatorial plane of the bare hemisphere.
In addition, simulations show two boojum defects located at the bubble's top and bottom tips.
Finally, the net force applied to the particle always remains positive in the numerical calculations, i.e. $F_{x}>0$: the colloid only self-propels toward its coated side (see Fig.\ref{fig:simulations}d).

% \subsection{Planar anchoring}
For planar confinement, Fig.\ref{fig:simulations}b and c, respectively, show the nematic field with their defects' configuration at low and high activation light intensities.
Independently of the illumination, a vertical Saturn ring is formed at the contact line between the isotropic bubble and the particle. 
In addition, simulations show the emergence of a boojum defect at the right tip of the bubble, confirming our experimental findings.
At low light intensity, the bubble's covering on the particle is minimal and we found a self-propulsion directed toward the coated side ($F_x>0$, see \Fref{fig:simulations}e).
At higher light intensity, the bubble's covering further increases and we found a self-propulsion directed away from the coated side ($F_x<0$, see \Fref{fig:simulations}f) in this scenario. 
% The $\Omega$ parameter identifies a normalized magnitude of the pressure force, distributed as different arrow sizes and colors. The parameter is defined as $\Omega = |\frac{\vec{F}_{sim}\cdot\vec{P}_s}{|\vec{F}_{sim}|^2}-1|$, with $\vec{F}_{sim}$ the computed force at each point on the surface of the particle-isotropic bubble composite, $P_s = P_p$ if located on the particle surface, and $P_s = P_b$ if located on the bubble surface.  
% These numerical results are consistent with our experimental observations.
\\
\\
Finally, simulations show that the magnitude $|F_x|$ of the self-propulsion force is higher with homeotropic confinement than with planar confinement; a discrepancy that is further confirmed experimentally. 
In the next section, we present a qualitative model describing the emergence of the colloid's motility. 
In particular, our model reveals how the bubble's covering (and therefore the light intensity) controls the direction of the motility in the case of planar confinement.

\section*{Self-propulsion mechanism}
\label{sec:simulations}
To characterize the self-propulsion mechanism, we analyze the force distribution on the surface of the particle-bubble system. 
The stress tensor $\bm{\sigma}$ exerted on the activated particle can be decomposed into three components as $\bm{\sigma} = P_p \mathbf{1}_{S_p}\mathds{1} + P_b \mathbf{1}_{S_b}\mathds{1} + \bm{\sigma}^{\textrm{defect}}$,
% \begin{align}
% \label{eq:pfd}
%     \sigma = P_p \mathbf{1}_{S_p}\mathds{1} + P_b \mathbf{1}_{S_b}\mathds{1} + \sigma^{\textrm{defect}}\;,
% \end{align}
% \begin{align}
% \label{eq:pfd}
%     \sigma=\int_{S_p}P_p\mathrm{d}\vec{S}+\int_{S_b}P_b\mathrm{d}\vec{S} + \int_{S_p \cup S_b}\sigma^{\textrm{defect}} \cdot d\vec{S}\;,
% \end{align}
where $\mathbf{1}$ and $\mathds{1}$ are, respectively, the indicator function and the identity. 
The first contribution represents the pressure exerted on $S_p$, the surface of the particle uncovered by the bubble. The second contribution similarly represents the pressure exerted on $S_b$, the surface of the bubble. 
Finally, the third contribution $\bm{\sigma}^{\textrm{defect}}$ is an additional inhomogeneous stress tensor arising from the presence of topological defects (Saturn rings and boojums).
Far from these defects, the force exerted on a surface $d\vec{S}$ of the system is therefore, respectively, $-P_p d\vec{S}$ on $S_p$ and $-P_b d\vec{S}$ on $S_b$.
We will further assume that the pressures $P_p$ and $P_b$ are constant and induced by splay deformations such that $P_p\propto K^p_1/R_p^2$ and $P_b\propto K^b_1/R_b^2$, with $K^p_1$, $R_p$ and $K^b_1$, $R_b$ being respectively the splay modulus and radius of the particle and bubble. 
Note that the splay modulus depends on the local director field, which decreases with temperature.
Since the bubble remains at a higher temperature than the surrounding LC, we therefore have $K^b_1 < K^p_1$.
% It is reasonable to consider $K^b_1 < K^p_1$ since the elastic modulus depends on the local scalar order parameter $S_0$, which decreases with temperature Because the bubble remains at a higher temperature than the surrounding LC, i (see Appendix B for further explanation of the numerical implementation). 
In addition, we experimentally observe that $R_b>R_p$, hence entailing $P_b<P_p$. 
The $x$ component of the total force exerted on the system is obtained by projecting $\int\bm{\sigma}\cdot d\vec{S}$ onto $\vec{e}_x$ and reads (see Methods)
\begin{align}
\label{eq:force_x_compo}
    F_x = F_x^P + F^{\textrm{defect}}_x = \pi h^2 \Delta P + F^{\textrm{defect}}_x\;,
\end{align}
where $\Delta P = (P_p-P_b) > 0$, $h$ is the radius of the contact circle between the bubble and the bead, and $F^{\textrm{defect}}_x$ is the $x$ component of the defect-generated force $\int \bm{\sigma}^{\textrm{defect}}\cdot d\vec{S}$ (see \Fref{fig:Angle_sketch}).

% \subsection{Homeotropic anchoring}
For homeotropic confinement, the Saturn ring is located in the equatorial plane $xy$ and remains localized on $S_p$. Therefore, it always pushes the colloid in the $x$ direction and $F^{\textrm{defect}}_x$ remains positive.
% , when following the convention of \Fref{fig:simulations}d and e.
Consequently, \eref{eq:force_x_compo} indicates that the particle always self-propels toward the coated hemisphere, regardless of the light intensity.

% \subsection{Planar anchoring}
For planar confinement, we use numerical simulations (see Methods) to show that
\begin{align}
    F^{\textrm{defect}}_x = -C\sin(2\theta_b)\;,
\end{align}\label{eq:Fdefect}
where $C$ is a positive constant and $\theta_b$ is the polar angle of the contact line between the bubble and the bead (see Figure \ref{fig:Angle_sketch} for a graphical definition of $\theta_b$).
Therefore, $F^{\textrm{defect}}_x$ changes sign at $\theta_b = \pi/2$.
When the illumination is weak, the bubble coverage is small, and $\theta_b>\pi/2$.
In this case, $F^{\textrm{defect}}_x$ remains positive and the colloid moves toward the coated hemisphere. 
When the illumination is intense, the bubble coverage is large, and $\theta_b<\pi/2$. 
In this case, $F^{\textrm{defect}}_x$ becomes negative. 
If $\theta_b<\arctan(\frac{C}{\pi R_p^2 \Delta P})$,  $F^{\textrm{defect}}_x$ exceeds the pressure forces and the particle propels away from the coated hemisphere.
Eq \eref{eq:force_x_compo} hence also describes the reversal of motility observed experimentally upon varying the light intensity when the confinement is planar.

% \begin{figure*}[h]
%     \centering
%     \includegraphics[width=0.9\linewidth]{manuscript/figs/fig3_2.pdf}
%     \caption{Geometrical configuration of the system particle-bubble mapped from the experimental observations. \textbf{a,} Schematic showing the definitions of $\theta$, $\theta_b$, $\theta_b'$, and $h$. The bubble's size and its relative position with respect to the particle are fitted from experimental images (see \textbf{b, c,} for planar confinement and \textbf{d,} for homeotropic confinement)}
%     \label{fig:Angle_sketch}
% \end{figure*}

\section*{Single-body active dynamics}
\label{sec:trajectories}
\begin{figure*}[h]
\centering
\includegraphics[width=0.95\linewidth]{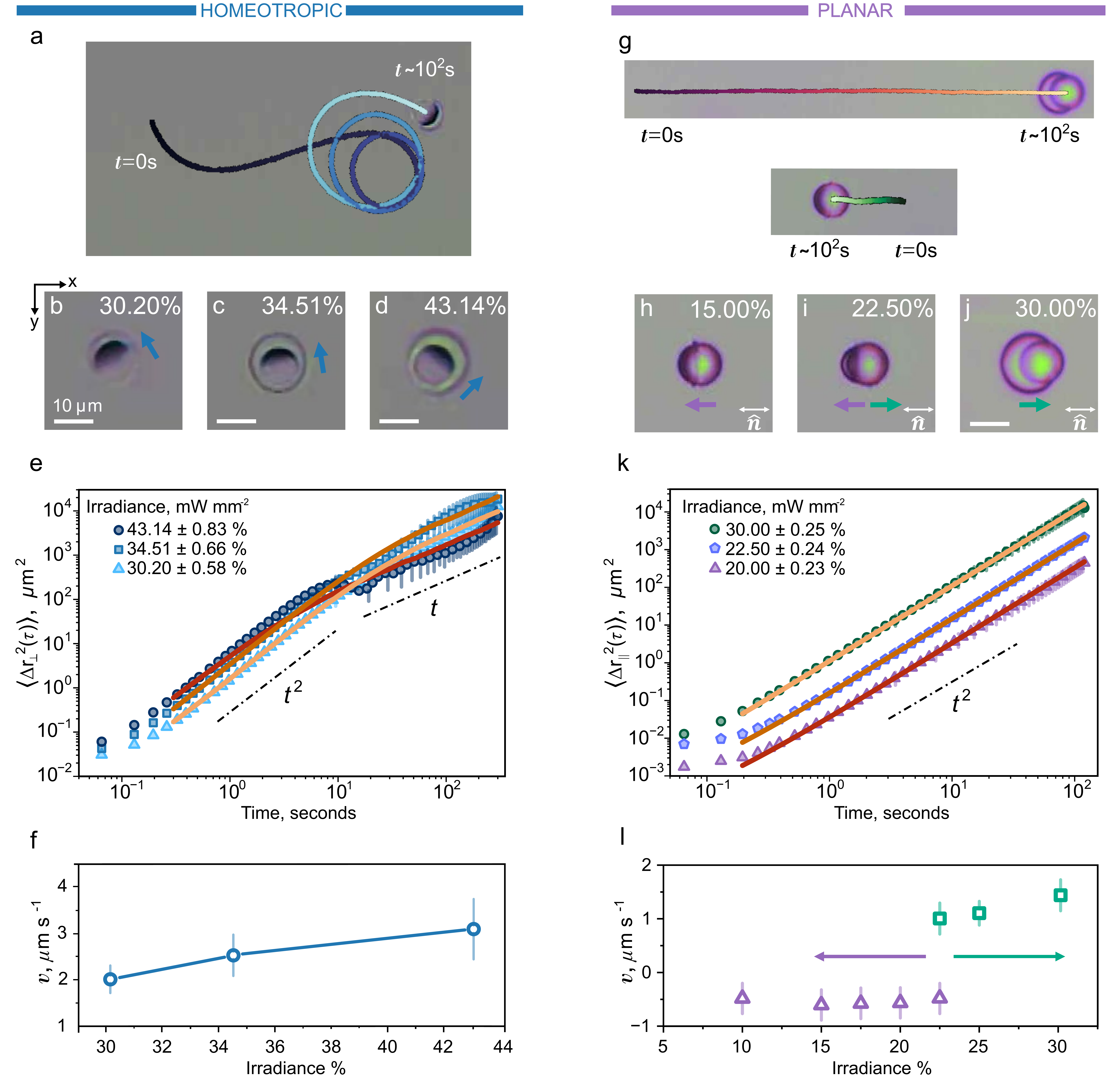}
\vspace{0.2cm}
\caption{Trajectory analysis of the light-activated Janus colloids for homeotropic (left) and planar (right) confinements. 
\textbf{a,g} Trajectories of representative active colloids. Lighter path colors indicate longer times. 
\textbf{b, c, d, h, i, j} Bright-field optical microscopy images for different irradiances with arrows indicating the direction of the self-propulsion. Note that the area covered by the isotropic phase increases with irradiance.
\textbf{e,k} Experimental mean-squared displacement of the colloids for different irradiances. 
Continuous lines represent the MSD of our theoretical models \eref{eq:ABP} and \eref{eq:theory_planar} with the parameters given in Table \ref{tab:DTV}. All irradiance values are given in units of mW mm$^{-2}$.
\textbf{f,l} Self-propulsion speed as a function of the light intensity percentages. 
For planar anchoring, negative (positive) values indicate a speed directed toward (away) the coated side.
}
\label{fig:trajectories}
\end{figure*}

We now describe the colloid's trajectory under various light intensities expressed as a percentage value of a maximum irradiance of $257.50\pm 0.50$ mW mm$^{-2}$.
% The self-propulsion of the Janus particle in the cell is directed along the polar angle $\phi$ in the $xy$ plane.
% \subsection{Homeotropic anchoring}
% In the case of an homeotropic confinement, we conducted experiments at three light intensities of 70$\%$, 80$\%$, and 100$\%$, respectively corresponding to a percentage value out of a maximum light power of $106.90\pm 0.01$ mW. To account for the illumination areas, the irradiance is a more useful parameter, which in the homeotropic case corrrespond to  $77.76\pm 1.49$, $88.87\pm 1.70$, and $111.09\pm 2.13$ mW mm$^{-2}$.
We start our analysis with the case of homeotropic confinement, which corresponds to the left column of \Fref{fig:trajectories} (for a complete trajectory, see \Fref{fig:trajectories}a) as well as to the supplementary movie Mov1. We conducted experiments at three irradiance values: $30.20\pm 0.58$ $\%$, $34.51\pm 0.66$ $\%$, and $43.14\pm 0.83$ $\%$.
% Note that the particle trajectories are random, since the LC bulk anchoring seems homogenous around the colloid, with the emergent asymmetry at the colloid level.
As highlighted in \Fref{fig:trajectories}b-d, both the bubble's radius and the colloid's self-propulsion increase with light intensity: as intuitively expected, symmetry breaking increases with illumination.
The experimental mean-square displacements (MSDs) for the three irradiances are displayed in \Fref{fig:trajectories} e. 
They all feature an intermediate-time ballistic regime where $\langle \vec{r}^2\rangle \propto t^2$ and a long-time diffusive regime where $\langle \vec{r}^2\rangle \propto t$ (details on the tracking methodology and MSD's determination can be found in the Methods).
% We further define the instantaneous speed of a particle as the stable value of $v=\sqrt{v_x^2+v_y^2}$ reached in steady-state, after the relaxation of the LC around the colloid. This definition stems from active particle models in which drag forces balance active forces, yielding a constant value of $v$\cite{lauga2020fluid}. 
\Fref{fig:trajectories}f shows the steady-state speed as a function of light intensity: consistently with numerical simulations, it increases monotically.

% \subsection{Planar anchoring.}

In the case of planar confinement, we carried out experiments at irradiances $20\pm 0.23$ $\%$, $22.5\pm 0.24$ $\%$, and $30\pm 0.25$ $\%$.
The results are summarized on the right side of \Fref{fig:trajectories}.
Typical trajectories of the colloids are displayed in \Fref{fig:trajectories}g; the particles are restricted to move in the direction of the nematic director. 
However, they can either self-propel toward (negative speed) or away from (positive speed) the coated side depending on the light intensity, as shown in \Fref{fig:trajectories}h-j and supplementary movies Mov2, Mov3 and Mov4.
% MSDs are reported in \Fref{fig:trajectories}k, where a clear monotonic increase occurs with the irradiance, as happens with the homeotropic case. Since the particle trajectories have larger persistances, the ballistic behavior remains even for the long-time regime.
% The illumination threshold of $57.94\pm 0.11$ mW mm$^{-2}$ marks the motility's reversal for non-monotonic speeds. 
The illumination threshold of $22.5\%$ marks the motility's reversal: at lower light intensities, $v$ remains constant around $\sim$$-0.5$ $\mu$m s$^{-1}$, while for higher intensities it monotonically increases to reach values close to 1.5 $\mu$m s$^{-1}$ (see \Fref{fig:trajectories}l).
MSDs are reported in \Fref{fig:trajectories}k for various irradiances: they all feature a long-time ballistic regime where $\langle \vec{r}^2\rangle \propto t^2$.

We are now set up to devise a theoretical model that accounts for these features.
%
% \section{Theoretical modelization}
% \label{sec:theory}
%
% We model the motion of an activated colloid in the $x-y$ through active matter dynamics.
The most prominent experimental signature that constrains modeling is the MSD.
Leveraging active matter dynamics, we build two minimal models compatible with this signature for each type of confinement. 

% \subsection{Homeotropic anchoring}
For homeotropic confinement, the direction of the self-propulsion of the particle remains unconstrained.
We thus model the colloid as an Active Brownian Particle (ABP) \cite{romanczuk2012active,solon2015active}, assuming that its position $\br(t)$ and orientation $\phi(t)$ follow. 
\begin{align}\label{eq:ABP}
\dot{\br} &= v \bu(\phi) + \sqrt{2T}\vec{\eta}, & \dot{\phi}&=\sqrt{2D}\:\xi\;,
\end{align}
where $v$ is the speed magnitude, $\phi$ ($\bu(\phi)$) is the polar angle (unit vector) of the velocity direction in the $xy$ plane, $T$ and $D$ controls the amplitude of translational and angular fluctuations, $\xi$ is a Gaussian white noise of unit variance, and $\vec{\eta}$ is a Gaussian vector with independent components of unit variance.
The MSD $\langle \Delta \br^2 \rangle$ of dynamics (\ref{eq:ABP}) has three distinct regimes; a diffusive one at small times, where $\langle \Delta \br^2 \rangle \sim 4Tt $, a ballistic regime at intermediate times where $\langle \Delta \br^2 \rangle \sim v^2t^2 $ and a diffusive regime at long times where $\langle \Delta \br^2 \rangle \sim 2\left(\frac{v^2}{D}+2T\right)t $. 
% These three regimes are related to $v$, $T$ and $D$ according to 
% % \begin{align}
% %     \langle \Delta \br^2 \rangle &\underset{t\to 0}{\sim} v^2t^2, & \langle \Delta \br^2 \rangle &\underset{t\to \infty}{\sim} \frac{v^2}{2T}t\;. \\
% %     & \langle \Delta \br^2 \rangle \underset{t\to \infty}{\sim} \frac{v^2}{2T}t\;. 
% % \end{align}
% \begin{align}
%     \langle \Delta \br^2 \rangle = \begin{cases}
% 4Tt & \text{if } t \to 0 \\
% v^2t^2 & \text{if } t \sim 1 \\
% 2\left(\frac{v^2}{D}+2T\right)t & \text{if } t \to \infty 
% \end{cases}
% \end{align}
Using these three regimes and our experimental determination of the MSD, we can therefore extract the values of $v$, $T$ and $D$ that best describe the colloid dynamics. 
We detail our fitting methodology in the Methods and report the corresponding results in Table \ref{tab:DTV}.
Finally, we simulated \eref{eq:ABP} and superimposed the resulting MSD on the experimental measurements in \Fref{fig:trajectories}e, proving the validity of \eqref{eq:ABP} with our fitted parameters.

% \subsection{Planar anchoring}
For planar confinement, the self-propulsion of the colloid is directed along the nematic director, which we assume to be in the $x$ direction.
In addition, the particle is also subject to surrounding thermal fluctuations, so that we model its dynamics as 
\begin{align}
\label{eq:theory_planar}
    \dot{\br} &= v\vec{e}_x + \sqrt{2T}\vec{\eta}\;,
\end{align}
where $T$ controls the magnitude of the fluctuations, $v$ is the self-propulsion speed, and $\vec{\eta}$ is a Gaussian vector with independent components of unit variance.
The MSD of dynamics \eref{eq:theory_planar} features two distinct regimes; a diffusive one at short times where $\langle \Delta \br^2 \rangle \sim 4Tt $ and a ballistic one at long times where $\langle \Delta \br^2 \rangle \sim v^2 t^2 $.
% These two regimes are related to $T$ and $v$ according to 
% \begin{align}
%     \langle \Delta \br^2 \rangle &\underset{t\to 0}{\sim} 4T t, & \langle \Delta \br^2 \rangle &\underset{t\to \infty}{\sim} v^2 t^2\;.
% \end{align}
% \begin{align}
%     \langle \Delta \br^2 \rangle = \begin{cases}
% 4Tt & \text{if } t \to 0 \\
% v^2 t^2 & \text{if } t \to \infty 
% \end{cases}
% \end{align}
Using our experimental determination of the MSD, we extract the values of $T$ and $v$. The corresponding results are reported in Table \ref{tab:DTV} in the Methods.
Finally, we simulated \eref{eq:theory_planar} and superimposed the MSD on top of its experimental counterpart in \Fref{fig:trajectories}k.

% \begin{figure}[h]
%     \centering
%     \includegraphics[width=0.9\columnwidth]{manuscript/figs/MSD_fit.pdf}
%     \caption{Mean squared displacement}
%     \label{fig:MSD_fit_abp}
% \end{figure}

\section*{Quorum sensing}
\label{sec:MIPS}
\begin{figure*}[]
\centering
\includegraphics[width=1\linewidth]{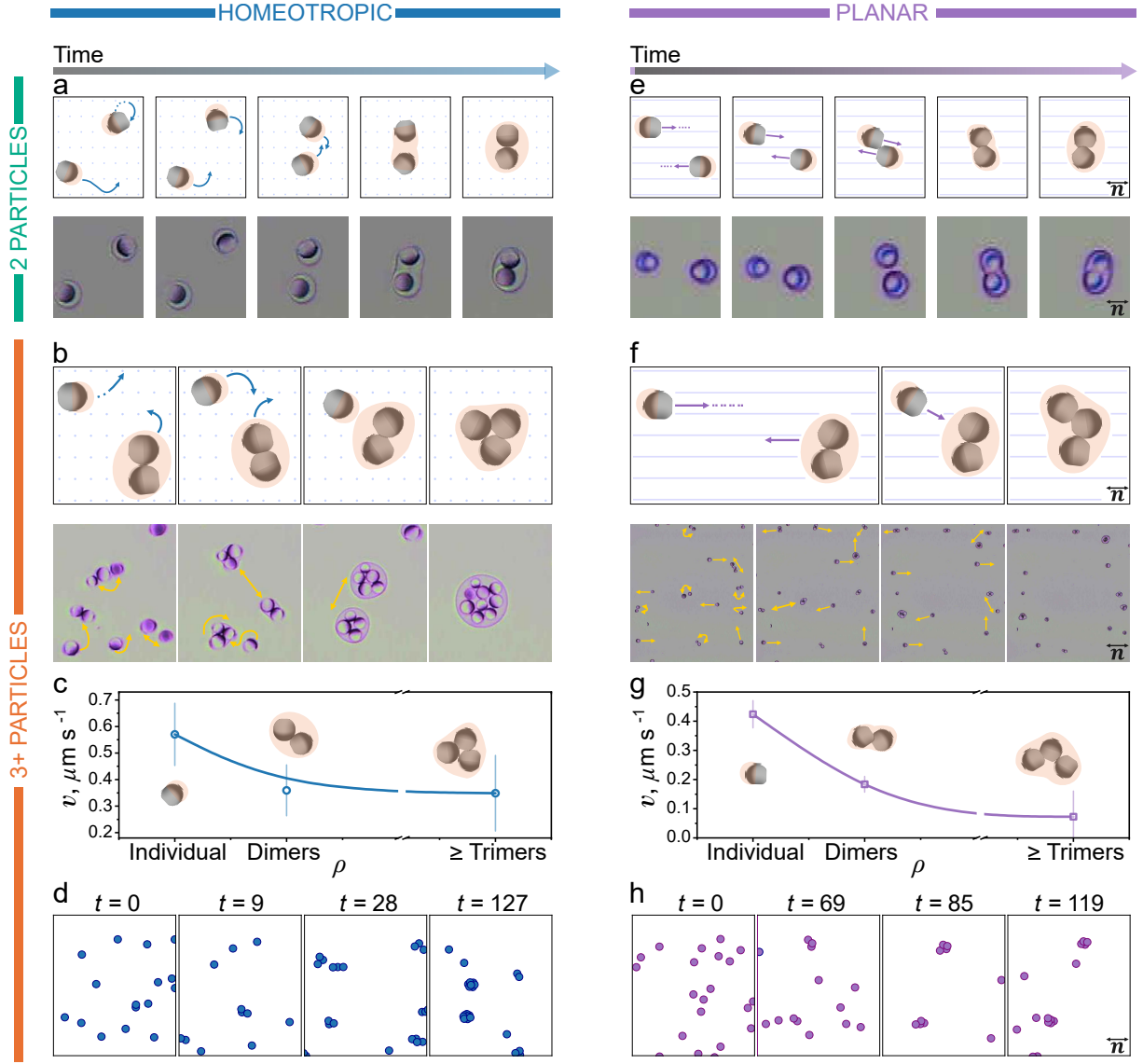}
\vspace{0.2cm}
\caption{
Many-body dynamics of the active colloids for homeotropic (left) and planar (right) confinements, starting with the two-body dynamics (first row) before the three-body dynamics (second row).
\textbf{a, e} Schematic of the two-body evolution (top subrow) with their corresponding experimental snapshots (bottom subrow).
\textbf{b, f} Schematic of the many-body evolution (top subrow) with their corresponding experimental snapshots (bottom subrow).
\textbf{c, g} Experimental self-propulsion speed of one-body, two-body and three-or-above-body clusters. Plain lines are fits (performed in log-log) to \eref{eq:QS_speed} with $\lambda=0.24$ (c) and $\lambda=0.88$ (g). 
\textbf{d, h} Numerical simulations of quorum-sensing active dynamics \eref{eq:mbody_MIPS_homeotropic} with $D=0.096$, $T=0.02$, $v=1.25$, $\lambda=0.24$ (d) and $D=0$, $T=0.01$, $v=1.0$, $\lambda=0.88$ (h). The time numbers indicate the time step for each snapshot.
}
\label{fig:experimental_many_body}
\end{figure*}

We now discuss the many-body dynamics of the light-activated Janus colloids.
As shown in \Fref{fig:experimental_many_body}, the collision of two particles provokes the merging of their isotropic bubbles.
At the end of this collision, the two Janus colloids form a cluster entirely surrounded by a larger isotropic bubble (see \Fref{fig:experimental_many_body}a and e).
As shown in \Fref{fig:experimental_many_body}b and f, surrounding single Janus colloids can further aggregate to this two-particle cluster through additional collisions. 
At the end of this aggregation process, the isotropic bubble ends up encompassing all the colloids once again: it \textit{swallows} the newly aggregated particle.
As the cluster aggregates more and more colloids, its surrounding isotropic bubble becomes increasingly more spherical. 
% For further observations of this phenomenon, we refer to the supplementary movies Mov5 (homeotropic confinement) and Mov6 (planar confinement). 
This loss in asymmetry results in a corresponding decrease in the motility of the cluster with its size, as observed in the plots of \Fref{fig:experimental_many_body}c and g. 
This many-body clustering scenario is qualitatively independent of the confinement type, be it homeotropic (see Mov5) or planar (see Mov6).
We thus report only quantitative differences between these two cases, such as the actual speed of dimers, trimers, and larger clusters. 

% \subsection{Quorum-sensing modelling}
We effectively model the many-body interactions by iterating two operations at each time step of the dynamics.
First, we construct the graph $\mathcal{G}$ whose nodes correspond to the positions of the particles and whose edges connect the particles $i$ and $j$ only if their distance satisfies $|\br_i-\br_j|<1$.
Second, we attribute a quorum-sensing, density-dependent motility $v(\rho)$ to each connected component $\mathcal{C}$ of the graph $\mathcal{G}$.
The latter is routinely done in the active matter community, where a sufficiently decreasing density-dependent motility is known to trigger Motility-Induced Phase Separation (MIPS).
In addition, inside each connected cluster $\mathcal{C}$, we model the dynamics of the particles with pairwise Hookean interactions. 
This prevents particles from leaving the cluster, while leaving the dynamics of its center of mass unaffected.
Following our measurements performed in \Fref{fig:experimental_many_body}, we assume that the speed of the clusters $v(\rho)$ is an exponentially decreasing function,
\begin{align}
\label{eq:QS_speed}
    v(\rho_{\mathcal{C}})=v\exp(-\lambda (\rho_{\mathcal{C}}-1))\;,
\end{align}
where $\rho_{\mathcal{C}}$ is the total number of particles present in the cluster $\mathcal{C}$.
In \eref{eq:QS_speed}, $v$ is the speed of a single activated colloid and $\lambda$ characterizes the decay of the cluster's speed with its density.
Although we already extracted the experimental value of $v$ (see Table \ref{tab:DTV}), we still need to evaluate $\lambda$.
To this aim, we fit the experimental values of the cluster's speeds (\Fref{fig:experimental_many_body}c and g) to an exponential decay with the density. 
The resulting values of $\lambda$ for the two confinement scenarios are reported in the caption of \Fref{fig:experimental_many_body}.
% Note that the values of $v$ and $\lambda$ depend on the type of confinement as well as on the light intensities, as already highlighted on \ref{}.
We now couple the quorum-sensing interactions \eref{eq:QS_speed} to the single-body models developed in Section \ref{sec:trajectories} to arrive at many-body descriptions for both planar and homeotropic confinements.

% \subsubsection{Homeotropic anchoring}
With homeotropic confinement, we assume that the center of mass $\br_{\mathcal{C}}$ of cluster $\mathcal{C}$ evolves according to
\begin{align}
\label{eq:mbody_MIPS_homeotropic}
	\dot{\br}_{\mathcal{C}} = v(\rho_{\mathcal{C}})\bu(\phi_{\mathcal{C}})+ \sqrt{2T}\vec{\eta}_{\mathcal{C}},\quad & \dot{\phi}_{\mathcal{C}}=\sqrt{2D}\xi_{\mathcal{C}}\;,
\end{align}
where $v(\rho_{\mathcal{C}})$ is given by \eref{eq:QS_speed} while $\vec{\eta}_{\mathcal{C}}$'s and $\xi_{\mathcal{C}}$ are Gaussian white noises such that $\langle\eta^\alpha_{\mathcal{C}}(s)\eta^\beta_{\mathcal{C}'}(t)\rangle=\delta_{\alpha\beta}\delta_{\mathcal{C}\mathcal{C}'}\delta(t-s)$ and $\langle\xi_{\mathcal{C}}(s)\xi_{\mathcal{C}'}(t)\rangle=\delta_{\mathcal{C}\mathcal{C}'}\delta(t-s)$.

% \subsubsection{Planar anchoring}
With planar confinement, the clusters are restricted to self-propel in the $\vec{e}_x$ direction. 
We therefore model the evolution of the center of mass $\br_{\mathcal{C}}$ according to
\begin{align}
\label{eq:mbody_MIPS_planar}
	\dot{\br}_{\mathcal{C}} = v(\rho_{\mathcal{C}})\vec{e}_x + \sqrt{2T}\vec{\eta}_{\mathcal{C}}\;,
\end{align}
where $\vec{\eta}_{\mathcal{C}}$ is a noise similar to that in \eref{eq:mbody_MIPS_homeotropic}.

% \subsection{Numerical simulations}
Using our experimental estimates for $\lambda$, $v$, $T$ and $D$ (see Table \ref{tab:DTV}), we performed numerical simulations of \eref{eq:mbody_MIPS_homeotropic} and \eref{eq:mbody_MIPS_planar}.
For both confinements, we found that the many-body dynamics leads to a clustering phenomenon similar to the one observed in the experimental setting, thereby confirming the plausibility of our theoretical descriptions (see the time evolution simulation snapshots in \Fref{fig:experimental_many_body}d and h, as well as the supplementary movies Mov7 and Mov8).

The clustering mechanism at play in our experiments and simulations is distinct from Motility-Induced Phase Separation since particles, once aggregated to a cluster, are precluded from leaving it: a feature not present in MIPS models. 
This difference entails noticeable deviations from MIPS' phenomenology, one of them being that the phase separation we observe does not completely coarsen into a single dense cluster coexisting with a dilute background.

% \begin{figure*}
% \centering
% \includegraphics[width=\textwidth]{manuscript/figs/ABP_coarsening.pdf}
% \caption{Coarsening of many-body quorum sensing ABPs evolving according to \eref{eq:mbody_MIPS} with experimentally-extracted parameters from Tables \ref{tab:v_T} and \ref{tab:lambdas}: $\lambda=0.47$, $v=1.88$ and $T=0.026$. Time increases from left to right and particles are initially dispatched at random position and orientation in the system. As the particles evolve, they form dense clusters coexisting with a dilute phase.}
% \label{fig:ABP_coarsening}
% \end{figure*}

\section*{Outlook} \label{conclusions}
We have introduced a versatile experimental platform in which, upon illumination, Janus particles can be induced to self-propel in a nematic liquid crystal.
% We have shown that this new self-propulsion mechanism stems from the interplay between the mesogens' anchoring condition around the bead and at the platform's surface.
We demonstrated that this new self-propulsion mechanism has two important consequences. 
First, it allows for an external control of the colloids' self-propulsion by adjusting the mesogens' anchoring at the surface of the platform.
Second, it mechanically endows the particles with quorum-sensing interactions, producing a local slowing down of motility with the local density.
% Taking into account both features, we proposed minimal theoretical models for both single-colloid and many-colloids dynamics. 
% The former provided a quantitative description of the experimental mean-square displacement while the latter highlighted the emergence of a Motility-Induced Phase Separation also observed in the experiments.
Our experimental platform paves the way toward a finer control of active assemblies endowed with life-like properties.
%An interesting avenue for further studies would be to enforce a spatial, time-dependent direction of the nematic director at the platform's surface. 
%As colloids are driven along the director field, this would create complex trajectories that will interplay with quorum-sensing interactions, potentially leading to novel collective behaviors.

\section*{Acknowledgements}
A. T.-V. thanks Prof. Erick Sarmiento-G\'{o}mez for the discussions and technical feedback on the experiments and data analysis, Prof. Teresa Lopez-Leon for fruitful discussions, Dr. Zhengyang Liu and Lars K\"{u}rten for the suggestions on the particle tracking methodology, and M.Des. Deydreth Martinez-Alvarez for the technical support of the figures. The authors gratefully acknowledge Prof. Stuart Rowan for sharing their polarized optical microscope, which allowed us to perform the experiments. This work used the Searle Cleanroom at the University of Chicago, funded by award number C06RR028629 from the National Institutes of Health - National Center for Research Resources. Instrumentation was purchased with generous funding provided by The Searle Funds at The Chicago Community Trust (Grant A2010-03222). %The SEM technique was performed at the Pritzker Nanofabrication Facility of the Institute for Molecular Engineering at the University of Chicago, which receives support from Soft and Hybrid Nanotechnology Experimental (SHyNE) Resource (NSF ECCS-2025633), a node of the National Science Foundation’s National Nanotechnology Coordinated Infrastructure, RRID: SCR\_022955. 
This work was also supported by the University of Chicago Materials Research Science and Engineering Center, which is funded by the National Science Foundation under award number DMR-2011854. 

\begin{figure*}[h]
\centering
\includegraphics[width=0.9\linewidth]{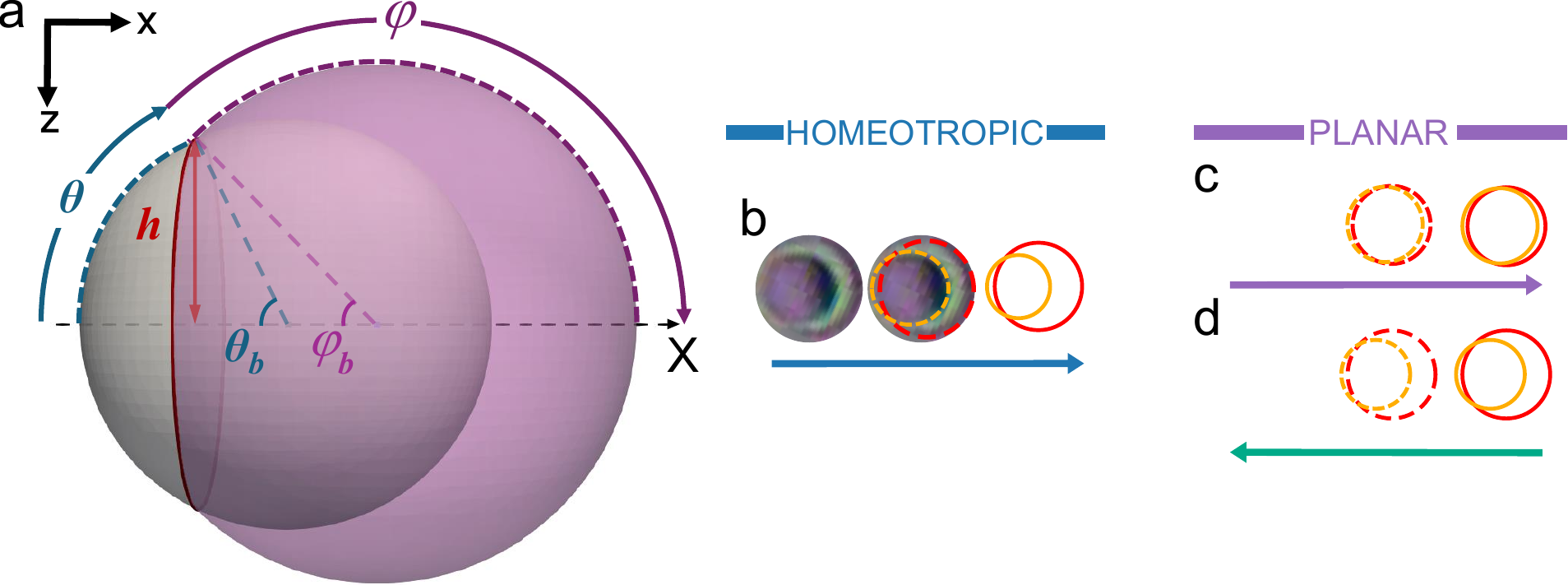}
%\renewcommand{\figurename}{Extended Data Fig.1}
%\captionsetup{name={Extended Data Fig.1}, labelformat=empty}%,labelsep=none}
\vspace{0.2cm}
\caption{Geometrical configuration extracted from experimental observations. \textbf{a} Schematic showing the definitions of $\theta$, $\varphi$, $\theta_b$, $\varphi_b$, and $h$. \textbf{b,c,d} Determination of the isotropic bubble and its relative position with respect to the particle center from experimental images for both homeotropic (\textbf{b}) and planar confinement (\textbf{c, d}).}
\label{fig:Angle_sketch}
\end{figure*}

\begin{figure*}[h]
\centering
\includegraphics[width=0.8\linewidth]{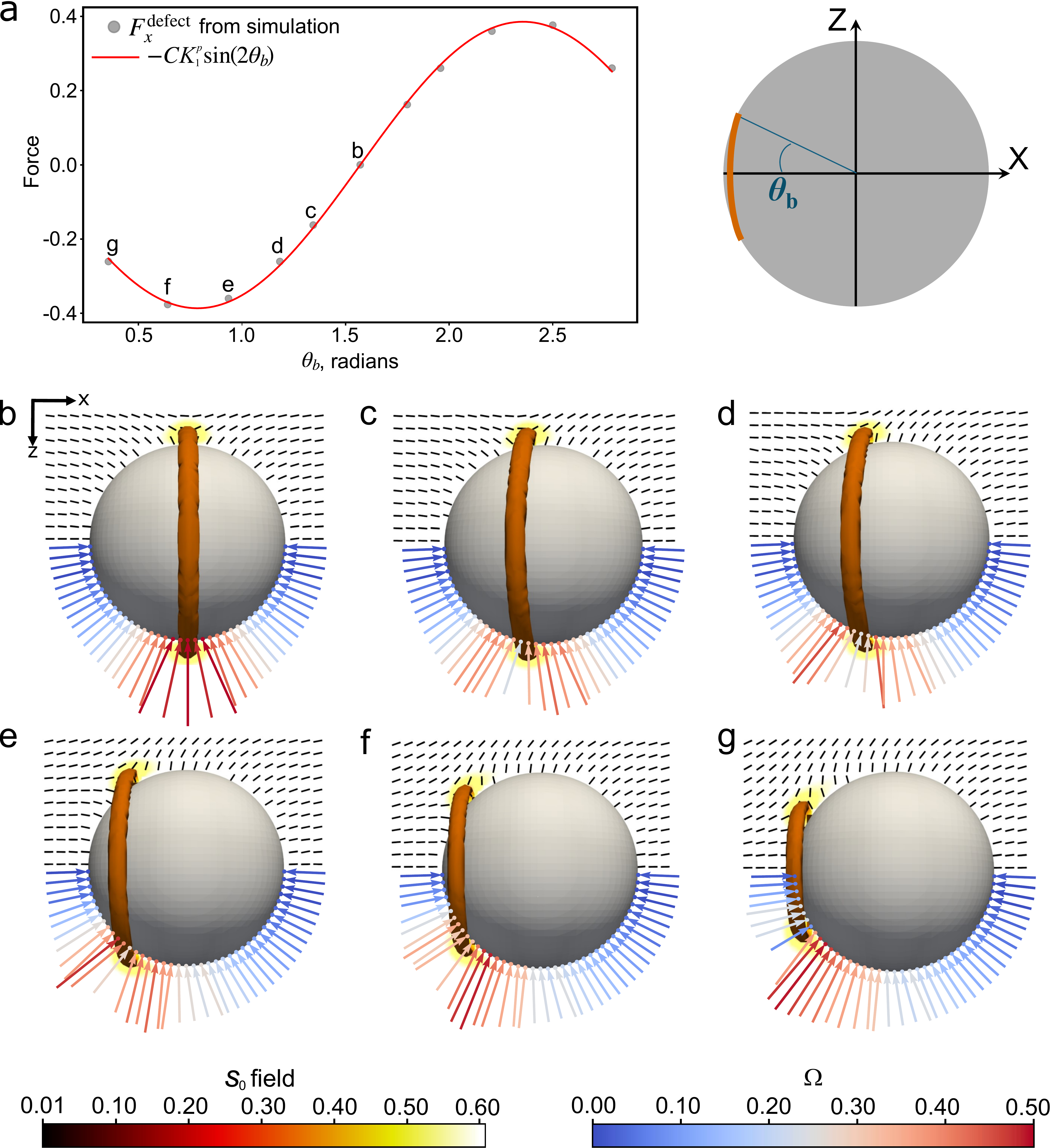}
\vspace{0.5cm}
\caption{Force distribution on a passive particle with planar confinement. \textbf{a} \({F}^\text{defect}_x\) as a function of the Saturn ring's position \(\theta_b\). 
Dots indicate numerical results while the plain line corresponds to theoretical prediction \eqref{eq:Fdefect}. \textbf{b--g} Numerical force distributions on the particle's surface for \(\theta_b = \pi/2\), 1.344, 1.183, 0.935, 0.641, and 0.355, respectively. The scalar nematic order parameter $S_0$ allows the visualization of the defects while the arrow colormap $\Omega$ indicates the magnitude of the force's component due to the presence of defects, namely $\bm{\sigma}^{\textrm{defect}}d\vec{S}$. It is defined as $\Omega = \big|\frac{\vec{F}_{sim}\cdot\vec{P}_p}{|\vec{F}_{sim}|^2}-1\big|$, with $\vec{F}_{sim}$ the numerically computed force and $P_p$ the pressure on the particle surface. Thus, $\Omega$ deviates from zero only when the local defects contribute to the force.}
\label{fig:SI_particle}
\end{figure*}

\section*{Methods}
\label{sec:methods}
\textbf{Experimental setup.}
Our Janus particles are $10$ $\mu$m silica spheres prepared according to standard protocols \cite{musevicOptExpress2010}, and half-coated with a 30 nm titanium layer. Their motility was constrained to 2D using $12$ $\mu$m spacers between the upper and lower glass cell walls. 
To induce homeotropic or planar anchoring of the liquid crystal, we treated the surfaces with DMOAP or brushed them with a microfiber
fabric after a polyvinyl alcohol (PVA) spin-coating. 
% The particles' surfaces were functionalized with DMOAP to induce homeotropic anchoring of the LC mesogens. To obtain homeotropic or planar anchoring of the LC bulk relative to the boundaries, we treated the glass walls with DMOAP or brushed them with a velvet fabric after a PVA spin-coating treatment, respectively.
The temperature of the confined LC was globally maintained at $35.40 ~\pm 0.01^\circ$C, in the nematic phase, by a dual top-bottom heating stage.
To activate the colloids, we used a LED source of nominal output power $500$ mW with a narrow spectrum line peaked at $\lambda = 542$ nm. 
% with a , and effective maximum output powers of $106.90\pm0.01$ mW and $41.20\pm0.01$ mW for the homeotropic and planar experiments respectively. 
We recorded particle trajectories at $15$ fps with a spatial resolution of 0.14006 $\mu$m/pixel at best. Therefore, the minimum measurable MSD accuracy is of the order of $\sim 10^{-2} ~\mu\text{m}^2$ and the minimum measurable \update{lag-time} is $\sim 0.065$ s.
Detailed information on every apparatus used and on the fabrication protocol can be found in the Appendix \ref{app:experimental_methods}.
\\
\\
\textbf{Experimental measurements.}
For tracking the Janus particles' center of mass, we followed standard protocols \cite{CUI2017452, Granick} and used ImageJ Fiji software \cite{Fiji} together with Trackpy \cite{Trackpy} for movie processing. 
The MSD is then computed as $\langle \Delta r^2 (\tau) \rangle:=
\frac{1}{n}\sum_{i=1}^n \Delta r^2_{i}(\tau)$
where $\Delta r^2_{i}(\tau):=[r(t_i+\tau)-r(t_i)]^2$, 
$n = N-\tau/\Delta t$, with $N$ the total number of measurements, $\Delta t$ the sampling frequency, $\tau$ the lag time and $t_i=i\Delta t$ ($i\in {0..N}$).
%To correctly estimate the statistical uncertainty of $\Delta r^2(\tau)$ without neglecting the correlations between the $\Delta r^2_{i}(\tau)$'s, we employed MUnCH \cite{cordoba2022munch,MUnCH}. 
\\
\\
\textbf{Numerical simulations.}
Our approach to modeling the LC surrounding the particle relies on a non-equilibrium, boundary-driven energy minimization. 
We focused on simulating the nematic phase of the LC and modeled the isotropic bubble as a fixed surface with planar anchoring, which is a good approximation of the experimental observations.
We assume the LC's total free energy $\mathcal{F}_{tot}$ to be given by
\begin{align}
    \mathcal{F}_{tot}=\int_V \left[f_{LdG}(\mathbf{Q})+f_{FR}(\mathbf{Q})\right]dV + \int_{\partial V} f_{s}(\mathbf{Q}) d\vec{S} \;,
\end{align}
where $\mathbf{Q}$ is the nematic traceless tensor, $f_{LdG}$ is the Landau-de Gennes energy, $f_{FR}$ the Franck--Riesler elastic energy and $f_s$ the surface anchoring energy ensuring the right boundary conditions for $\mathbf{Q}$. 
To model light activation, we assume that $\mathcal{F}_{tot}$ is further coupled to a spatially varying temperature field $T$, which disrupts the orientation of the mesogens.
As $f_{LdG}$ controls the nematic-isotropic transition through the parameter $U$, a natural way to enforce the coupling to temperature is to consider a $T$-dependent $U$ (see details in Appendix \ref{app:LC_numerics}). 
% We approximated $U$ as a linear function of $T$.
In the bulk, considering an isotropic homogeneous thermal conductivity, the steady-state temperature and nematic tensor are given by
\begin{align}
\label{eq:bulk_nematic}
    \nabla^2 T=&0\;, & \frac{\delta f_{LdG}}{\delta \mathbf{Q}} + \frac{\delta f_{FR}}{\delta \mathbf{Q}} =&0\;.
\end{align}
We further need to specify the boundary conditions for $T$ and $\mathbf{Q}$; at the contact surface with the isotropic bubble, they obey
\begin{align}
    T=&T_h, & \bn\cdot\frac{\partial f_{FR} }{\partial \nabla \mathbf{Q}}-\frac{\partial f_s}{\partial \mathbf{Q}}=&0\;,
\end{align}
where $T_h$ is a fixed warm temperature, $\bn$ is the normal unit vector on the surface and $f_s$ ensures planar anchoring. Likewise, at the boundary with the Janus particle, the fields respect
\begin{align}
    T=&T_c, & \bn\cdot\frac{\partial f_{FR} }{\partial \nabla \mathbf{Q}}-\frac{\partial f_s}{\partial \mathbf{Q}}=&0\;,
\end{align}
where $T_c$ is a fixed cold ambient temperature and $f_s$ ensures homeotropic anchoring (see details in Appendix \ref{app:LC_numerics}).
Finally, we enforced periodic boundaries in the $\bm{x}$ and $\bm{y}$ directions while distinguishing the two different scenarios for the cell boundaries in the $\bm{z}$ direction: (i) planar and (ii) homeotropic anchoring.
After finding $\mathbf{Q}$ and $T$, the force $\mathbf{F}$ exerted on each surface point of the compound particle--bubble is obtained as ~\cite{PhysRevE.58.7475,MouZhangInPerp}
\begin{equation}
F_j = \left( -\frac{\partial f_{FR}}{\partial\left( \partial_k Q_{pq}\right)} \partial_j Q_{pq} + (f_{FR} + f_{\text{LdG}}) \delta_{jk} \right) \nu_k\;,
\end{equation}
where summation over repeated indices is implied. $\nu_k$ is the k-th unitary vector normal to the particle-bubble surface. 
Summing this formula over the surface of the compound then yields the total self-propulsion force $F_x$.
Note that in our simulations we approximated the bubble's surface to be a spherical hull whose radius we retrieved in experimental pictures of the compound (see \Fref{fig:Angle_sketch}b-d).
\\
\\
\textbf{Self-propulsion force.}
Due to rotational symmetry around the $x$ axis, the pressure component of the self-propulsion force is directed along $\vec{e}_x$. Its amplitude $F_x^P$ is given by
\begin{equation}
  \begin{aligned}
F_x^P=\int_{S_p}P_p\mathrm{d}\vec{S}+\int_{S_b}P_b\mathrm{d}\vec{S},
    \end{aligned}
\end{equation}
where $S_p$ is the bare surface of the particle and $S_b$ represents the isotropic bubble surface. 
To simplify the derivation of $F_x^P$, we introduce in \Fref{fig:Angle_sketch} two spherical coordinate systems with distinct origins; one is centered on the particle while the other is centered on the isotropic bubble such that $S_p=\{(R_p, \theta,\psi), (\theta,\psi) \in [0,\pi]\times [0,2\pi]\}$ in the former and $S_b=\{(R_b, \varphi, \psi), (\varphi,\psi) \in [0,\pi] \times [0,2\pi] \}$ in the latter. 
We further define $\theta_b$ and $\varphi_b$ as the polar angles of the contact line between the particle and the bubble in both polar coordinate systems, respectively (see \Fref{fig:Angle_sketch}a). 
Using these definitions, we obtain the following:
\begin{equation}
\begin{aligned}
F_x^P &=
\int_{0}^{\theta_b} \pi P_p  R_p^2 \sin{2\theta}\mathrm{d}\theta +\int_{\varphi_b}^{\pi}\pi P_b R_b^2 \sin{2\varphi}\mathrm{d}\varphi\;,
\end{aligned}
\end{equation}
which can be further simplified into 
\begin{align}
\label{eq:force_x_pressure}
    F_x^P = \pi h^2 (P_p-P_b) \;,
\end{align}
where $R_p\sin\theta_b=R_b\sin\varphi_b = h$ is the radius of the contact line between the bubble and the particle.

We now determine the additional force exerted by the LC topological defects on the particle in the case of planar confinement.
In the absence of light, the defect ring of the particle is located in the $yz$ plane corresponding to $\theta=\pi/2$ and does not generate any net self-propulsion force. 
% However, in addition to the splay deformation, bend deformation also contributes to the elastic force near the Saturn ring defect, where the direction
% of the elastic force deviates from the particle’s surface normal direction, as shown in \Fref{fig:SI_particle}b and identified by the color code of the pressure force arrows ($\Omega$ parameter). 
However, if this ring is displaced away from $\theta_b=\pi/2$, we intuitively expect the emergence of a net force that tends to push back the particle in the
rest configuration corresponding to $\theta_b = \pi/2$.
% the symmetric elastic force is broken, and a net force emerges that tends to push the particle back to the rest configuration (see \Fref{fig:SI_particle}c-f). 
In addition, $\theta_b=0$ or $\pi$ correspond to hedgehog defects: These are stable configurations for which we expect $F^\text{defect}_{x}=0$.
The most simple $\pi$ periodic function that is odd with respect to $\pi/2$ and then vanishes at $\theta_b=0 \bmod \pi$ is $\sin(2\theta_b)$.
We therefore infer \eqref{eq:Fdefect},
%\begin{equation}
%\label{eq:fdefect_assum}
%F^\text{defect}_{x} = -CK_1^p %\sin({2\theta_b}),
%\end{equation}
where $C$ is a constant. Our intuitive expression \eqref{eq:Fdefect} agrees nearly perfectly with the result of our simulations (see \Fref{fig:SI_particle}). 
Additional simulation details are provided in Appendix \ref{app:LC_numerics}.
\\
\\
\textbf{Fitting parameters.}
To fit $T$, $v$ and $D$, we proceed in three steps for each intensity of light. 
Using $\langle\Delta \vec{r}^2(t)\rangle$ as a function of time on the logarithmic scale, we isolate the time range where it exhibits ballistic behavior. 
Fitting this time range to $v^2t^2$ allows us to retrieve the particle's speed $v$.
Focusing on smaller time scales, we then fit the mean square displacement to $4Tt+v^2t^2$, hence obtaining the translational diffusion $T$.
Finally, on the largest time scales, we fit the MSD to $(4T+2v^2/D)t$, thereby retrieving the angular diffusion $D$. 
The resulting fitted parameters are shown in Table \ref{tab:DTV}.  

\begin{table}[hb]
\def\colwid{2.7cm}
\def\colwidsec{1cm}
\hspace*{-0.3cm}
\begin{tabular}{ |>{\centering\arraybackslash}p{\colwid}|>{\centering\arraybackslash}p{\colwidsec}|>{\centering\arraybackslash}p{\colwidsec}|>{\centering\arraybackslash}p{\colwidsec}|}
 % \cline{2-5}
 %  \multicolumn{1}{c|}{} & \multicolumn{4}{c|}{$\langle \Delta \br^2\rangle \propto a t^{\alpha}$} \\
 %  \cline{2-5}
 % \multicolumn{1}{c|}{} & \multicolumn{2}{c|}{$t\to 0$} & \multicolumn{2}{c|}{$t\to\infty$} \\
 \hline
 Irradiance \bf{mW mm$^{-2}$} & $v$ & $T$ & $D$ \\ \hline
 \multicolumn{4}{|c|}{\textit{homeotropic confinement}} \\ \hline 
 $30.20\pm0.58$ $\%$ & 1.20 & 0.019 & 0.096  \\ \hline
 $34.51\pm0.66$ $\%$ & 1.84 & 0.026 & 0.095  \\ \hline
 $43.14\pm0.83$ $\%$ & 2.45 & 0.067 & 0.705 \\ \hline
 \multicolumn{4}{|c|}{\textit{planar confinement}} \\ \hline 
 $20.00\pm0.23$ $\%$ & 0.17 & 0.0014 & -  \\ \hline
 $22.50\pm0.24$ $\%$ & 0.69 & 0.0026 & -  \\ \hline
 $30.00\pm0.25$ $\%$ & 1.03  & 0.010 & - \\ \hline
\end{tabular}
\caption{Values of $v$, $T$ and $D$ fitted from the experimental MSD. Note that we set $D=0$ for planar confinement according to \eref{eq:theory_planar}.}
\label{tab:DTV}
\end{table}

\onecolumn
\begin{appendices}
\section{Experimental methods}
\label{app:experimental_methods}
\noindent\textbf{Sample preparation.}
Janus particles were prepared by standard methods \cite{musevicOptExpress2010} using $10$ $\mu$m silica spheres functionalized with amine groups (CD Bioparticles, NY USA). A 30 nm layer of titanium was deposited on top of the particles (AJA ATC-Orion 8E e-beam evaporation system, AJA International Inc, MA USA), verified by electron microscopy (FEI Quanta 650 FEG SEM, Thermo Fisher Scientific Inc, USA). Afterward, to induce LC surface homeotropic anchoring on the Janus particles, they are suspended in a $2\%$ deionized water solution ($18.2$ M$\Omega$ cm, Millipore, Bedford, MA USA) of dimethyloctadecyl[3-(trimethoxysilyl)propyl]ammnonium chloride (DMOAP) (Sigma-Aldrich, USA) for 15 min, then rinsed profusely with deionized water, and finally baked at 100$^\circ$C for one hour. 
%It is well known that DMOAP chemically functionalizes silica particles after a hydrolysis reaction of the silane molecules that bond to the oxygen-silica molecules. The high-temperature curing process permits the polymerization of the silane monomers to form polysiloxanes. However, in our case, amine groups (NH$_2$) do not disturb silanization. On the contrary, it has been shown that these groups may improve the chemical reactions for various silane molecules \cite{polymers2018}. The capped side of the Janus particles is also silanized, corroborated by previous tests done on TiO$_2$ surfaces \cite{RSCAdvances2020}. 
The dried Janus particles are mixed at a dilute concentration of 0.01 wt$\%$ in 4-Cyano-4'-pentylbiphenyl (5CB) LC, purchased from Hebei Maison Chemical Co., Ltd (China). Thermogravimetric measurements were performed with a TGA instrument (TA Instruments, USA) to ensure the final particle concentration. We built cells to confine the LC-particle suspension for experimental observations using soda-lime glass microscope slides (Thermo Scientific, NH USA). \UpdatE{To induce homeotropic anchoring on the glass cell, we cleaned the glass slides with a half-half mixture of ethanol and acetone, then rinsed} with deionized water and air plasma treatment to eliminate any trace of organic pollutants. Later, the glass slides were also treated with DMOAP. \UpdatE{To induce planar anchoring, we cleaned the glass slides with piranha solution ($70\%$ H$_2$SO$_4$ ($95.0\%$-$98.0\%$ pure, Sigma-Aldrich, USA), $30\%$ H$_2$O$_2$ ($30\%$ dilution, Fisher Scientific, PA USA)) and then rinsed profusely with deionized water. The glass slides were then dried with nitrogen. Subsequently, we spin-coated each slide for 90 s with an aqueous solution of polyvinyl alcohol (PVA) at 10 wt$\%$ and a molecular weight of 13 kDa (Sigma-Aldrich, USA). After spin coating, the slides were baked for one hour at 100$^\circ$C. Then, we brushed each coated slide in one direction with a clean velvet fabric to create micro-grooves. The adhesiveness of the polymer alongside the microgrooves that are etched onto the polymer yields the planar anchoring of the LC mesogens. After brushing, we oriented the top and bottom glass pieces of the cell so that their brushing was in opposite directions to ensure homogeneity. For both homeotropic and planar anchoring cases,} we used 12 $\mu$m thick mylar films as cell spacers (Premier Lab Supply, FL USA), and 5-minute epoxy resin to seal cells (Devcon, MA USA). Slight variations in the cell thickness are possible due to the method of assembly. The LC colloidal suspension was then heated to 50$^\circ$C and gently introduced into the glass cells by capillary effects. Cells were immediately sealed with 5-minute epoxy resin and removed from the hot plate to quench the system; otherwise, Janus particles would reorganize and cluster because they are prone to being dragged by the nematic-isotropic (NI) interface. The LC was allowed to relax to obtain perfect homeotropic \UpdatE{or parallel alignment.} 
\\
\\
\noindent\textbf{Light activation and microscopy.}
We use the AURA III LED engine (Lumencor, OR USA) as the activation source (bandpass filter $\lambda$ = 542 nm, FWHM = 33 nm. Maximum nominal output power = 500 mW). Light-power measurements are performed using a Thorlabs PM160T power meter (USA). The light source is assembled into a Leica DM-2700P polarizing microscope (Germany) in reflection mode using a liquid light guide and a collimator. We use a custom-made filter cube (Chroma Technologies, USA; dichroic mirror $\%T \geq 95\%$ for 470 nm $\leq \lambda \leq$ 480 nm, $\%R \geq 95\%$ for 495 nm $\leq \lambda \leq$ 545 nm, and transmission filter $\%T \geq 95\%$ for 445 nm $\leq \lambda \leq$ 470 nm and 605 nm $\leq \lambda \leq$ 650 nm). The collimated light beam passes through an N-Plan achromatic objective of $20\times$ magnification (\UpdatE{for the homeotropic anchoring experiments) or and objective of $50\times$ magnification (for the planar anchoring experiments)}. Simultaneously, a white LED source is used in transmission mode to image the sample. Movies are recorded with a Leica MC170 HD camera. Bright-field and polarized movies are recorded at 15.3735 fps, and the camera's spatial resolution is 0.34662 $\mu$m/pixel when using the 20$\times$ objective, and 0.14006 $\mu$m/pixel when using the 50$\times$ objective. Therefore, the minimum MSD that can be measured is of the order of $\sim 10^{-2} ~\mu\text{m}^2$ and the minimum measurable \update{lag-time} is $\sim 0.065$ s.
\\
\\
\noindent\textbf{Temperature control.}
We use a dual top-bottom heating stage HCS402 and a high-precision temperature controller mk2000 (resolution 0.001$^\circ$C; INSTEC, USA) to maintain samples at constant $35.40\pm 0.01^\circ$C in the nematic phase.
\\
\\
\noindent\textbf{Particle tracking analysis.}
ImageJ Fiji distribution software \cite{Fiji} was used for movie processing. For proper tracking of the center of mass of Janus particles \UpdatE{in the case of cells with homeotropic anchoring}, standard protocols were followed and can be found elsewhere \cite{CUI2017452, Granick}. Once the coordinates of the center of mass of the particles were computed, the trajectory paths were displayed using Trackpy \cite{Trackpy}. \UpdatE{For the planar anchoring case, the particle's tracking analysis was performed using the ImageJ TrackMate plugin \cite{trackmate}.} 
\\
\\
\noindent\textbf{Mean-squared displacement computation.}
To obtain the MSD from the trajectories of Janus particles, the $\gamma$-position 
of the particle as a function of time, $r_{b,\gamma}(t)$, is tracked.
Where $t=i \Delta t$, $1/\Delta t$ is the sampling frequency of the 
particle tracking technique, $N$ is the total number of measurements 
and $i=0,1,2,3...,N$. The MSD is then computed as,
$\langle \Delta r^2_\gamma (\tau) \rangle:=
\frac{1}{n}\sum_{i=1}^n \Delta r^2_{i,\gamma}(\tau)$
where $\Delta r^2_{i,\gamma}(\tau):=[r_{b,\gamma}(t+\tau)-r_{b,\gamma}(t)]^2$,
$n = N-\tau/\Delta t$, $\tau$ is the \update{lag-time} and
the sub-index $\gamma$ indicates the 
spatial direction of the measurement, $\gamma$ can be either 
parallel, $\parallel$, or perpendicular, $\perp$, to the nematic director, $\hat{\bm{n}}$. 
Experiments with a homeotropic anchoring at the walls of the 
glass cell allow measuring  $\langle \Delta r^2_{\perp} (\tau) \rangle$ 
while experiments with planar anchoring at the walls of the glass cell  
allow measuring  $\langle \Delta r^2_{\parallel} (\tau) \rangle$.
%To calculate $\langle \Delta r^2_\gamma (\tau) \rangle$ and 
%estimate its statistical uncertainty, $\sigma(\langle \Delta r^2_\gamma (\tau)\rangle)$,
%here we employ MUnCH \cite{cordoba2022munch,MUnCH}. By using 
%repeated block transformations, MUnCH can correctly estimate 
%the statistical error of any autocorrelation at any given \update{lag-time}.
%A common omission in the calculation of $\sigma(\langle \Delta r^2_\gamma (\tau)\rangle)$
%is to neglect the correlations inherent in the bead position data. 
%These correlations are important in viscoelastic materials,
%the uncertainty can be underestimated by a factor of 
%up to $20$ if the correlation in the bead position data is neglected \cite{cordoba2022munch}.

\section{Numerical implementation}
\label{app:LC_numerics}
A nematic LC can be described by a tensorial order parameter ${\bf Q}$. For a uniaxial nematic LC, ${\bf{Q}} = S_0({\bf{nn}} - {\bf{I}}/3)$, where the unit vector $\bf{n}$ represents the nematic director field, $S$ is the scalar order parameter of the nematic LC, and ${\bf I}$ is the identity tensor. The total free energy of a nematic LC comprises the Landau--de Gennes short--range free energy density $f^{\text {LdG }}$, the long-range elastic energy density $f^{\text {el }}$, and the anchoring-induced surface free energy density $f^{\text {surf}}$
\begin{equation}
\mathcal{F}_{\text {total }}=\int_V\left(f^{\text {LdG }}+f^{\text {el }}\right) \text{d} V+\int_{\partial V} f^{\text {surf }} \text{d} S,
\end{equation}
where $V$ and $\partial V$ are the volume and surface of the region of interest. The Landau--de Gennes free energy $f^{\text {LdG }}$ is given by
\begin{equation}
\label{LdG}
\begin{aligned}
f^{\mathrm{LdG}} &=\frac{A_0}{2}\left(1-\frac{U}{3}\right) \operatorname{Tr}\left(\mathbf{Q}^2\right)-\frac{A_0 U}{3} \operatorname{Tr}\left(\mathbf{Q}^3\right)\\
&+\frac{A_0 U}{4}\left(\operatorname{Tr}\left(\mathbf{Q}^2\right)\right)^2,
\end{aligned}
\end{equation}
where the parameter $A_0$ is the energy density scale and $U$ controls the equilibrium magnitude of $S_0$ through 
\begin{equation}
\label{Ueq}
S_0=\frac{1}{4}+\frac{3}{4} \sqrt{1-\frac{8}{3 U}}.
\end{equation}

The elastic free energy $f^{\text{el}}$ can be expressed using the Einstein summation notation as
\begin{equation}
\begin{aligned}
f^{e l}= & \frac{1}{2} L_1\left(\partial_k Q_{i j}\right)\left(\partial_k Q_{i j}\right)+\frac{1}{2} L_2\left(\partial_k Q_{j k}\right)\left(\partial_l Q_{j l}\right) \\
& +\frac{1}{2} L_3 Q_{i j}\left(\partial_i Q_{k l}\right)\left(\partial_j Q_{k l}\right)+\frac{1}{2} L_4\left(\partial_l Q_{j k}\right)\left(\partial_k Q_{j l}\right) .
\end{aligned}
\end{equation}

Here, the coefficients $L_1$, $L_2$, $L_3$, and $L_4$ can be mapped to the Frank elastic constant $K_1, K_2, K_3,$ and $K_{24}$, which are the splay, twist, bend, and saddle-splay modulus, respectively, through 
\begin{equation}
\begin{aligned}
L_1 & =\frac{1}{2 S_0^2}\left[K_2+\frac{1}{3}\left(K_3-K_1\right)\right], \\
L_2 & =\frac{1}{S_0^2}\left(K_1-K_{24}\right), \\
L_3 & =\frac{1}{2 S_0^3}\left(K_3-K_1\right), \\
L_4 & =\frac{1}{S_0^2} (K_{24}-K_{2}) .
\end{aligned}
\end{equation}
In our simulations, we set $L_1 = 1$ and $L_2 = L_3 = L_4 = 0$, equivalent to $K_1 = K_2 = K_3 = K_{24} = 0.758$, and $A_0=1$.
The surface anchoring free energy $f^{\text {surf}}$ is given by
\begin{equation}
f^{\text {surf}}=\frac{1}{2} W\left(\mathbf{Q}-\mathbf{Q}_{\text {surf }}\right)^2.
\end{equation}
Here, $\mathbf{Q}_{\text {surf }}$ is the preferred field of the surface, and $W$ is the anchoring strength.

To obtain a stable solution of the nematic LC, we minimize the total free energy. We define a molecular field as
\begin{equation}
\mathbf{H}=-\left[\frac{\delta \mathcal{F}_{\text {total }}}{\delta \mathbf{Q}}\right]^{s t},
\end{equation}
where $[...]^{st}$ is a symmetric and traceless operator. Assuming that all transitions are quasistatic processes, the evolution of the $\mathbf{Q}$-tensor for bulk points follows
\begin{equation}
\label{QBulk}
\partial_t \mathbf{Q}=\Gamma_s \mathbf{H},
\end{equation}
where $\Gamma_s=\Gamma / \xi_n$ is the relaxation constant.  Here, $\Gamma$ is related to the rotational viscosity $\gamma_1$ of LC by $\Gamma = 2 S_0^2/\gamma_1$ and $\xi_n$ is the nematic coherence length. For surface points, the evolution of the $\mathbf{Q}$-tensor is governed by
\begin{equation}
\label{QSurf}
\frac{\partial \mathbf{Q}}{\partial t}=-\Gamma_s\left[\bm{\nu} \cdot \frac{\partial \mathcal{F}}{\partial \nabla \mathbf{Q}}+\left\{\frac{\partial f_{\text {surf }}}{\partial \mathbf{Q}}\right\}^{s t}\right],
\end{equation}
where the unit vector $\bm{\nu}$ represents the surface normal (pointing outside of the LC domain).

In the simulation, we define the bubble and particle regions according to the experimental images.
On the bubble surface, a degenerate planar anchoring condition is assumed. A homeotropic anchoring condition is imposed on the surface of the particle with $W = 1$. Because the LC is in the isotropic phase inside the bubble region (denoted as $\Omega$), we therefore only evolve the $\mathbf{Q}$-tensor outside the particle and bubble region using the above numerical approach.

We assume that the parameter $U$ in (\ref{LdG}) is linearly dependent on the local temperature $T$ according to \(U = 3.5 - 0.325 \times T\). The temperature field $T$ satisfies the Poisson equation $\frac{{\partial T}}{{\partial t}} = \nabla \cdot ({\bf \kappa}\cdot \nabla T)$, where ${\bf \kappa}$ is the thermal conductance tensor~\cite{PhysRevE.49.545}. For simplicity, we only consider an isotropic conductivity tensor ${\bf \kappa}=\kappa_0 {\bf I}$ with $\kappa_0=1$ in the simulation. The temperature on the surface of the unheated particle and the boundaries of the solution domain are set at 0, while the temperature on the heated bubble surface is set at \(T=T_{\text{cut-off}}\). In our discretized simulation, we update the temperature field from step $n$ to step $n+1$ according to
\begin{equation}
\label{Poisson}
\begin{aligned}
T_{i,j,k}^{n + 1} = & T_{i,j,k}^n + \Delta t \cdot \left( \frac{T_{i+1,j,k}^n + T_{i-1,j,k}^n + T_{i,j+1,k}^n }{h^2} \right. \\
& \left. + \frac{T_{i,j-1,k}^n + T_{i,j,k+1}^n + T_{i,j,k-1}^n - 6T_{i,j,k}^n}{h^2} \right), \\
& {\left. T \right|_{\partial {\Omega _{{\text{unheated particle}}}}}} = {\left. T \right|_{\text{boundary}}} = 0, \\
& {\left. T \right|_{\partial {\Omega _{{\text{bubble}}}}}} = T_{\text{cut-off}}.
\end{aligned}
\end{equation}
where $i$, $j$, and $k$ denote spatial coordinates, and $h$ denotes the unit size of the mesh in the simulation. From (\ref{Ueq}), we know that when \(U = \frac{8}{3}\), the system transitions to the isotropic phase, corresponding to a transition temperature of \(T_{\text{trans}} = 2.564\) in the simulation units. After bubble formation, its interior becomes isotropic, making \(T_{\text{cut-off}} \approx T_{\text{trans}}\). We set \(T_{\text{cutoff}} = 2\), slightly below \(T_{\text{trans}}\).

After obtaining a stable solution for the molecular field, the force $\mathbf{F}$ exerted by the Ericksen stress on each surface point of the particle and bubble can be calculated using the following equations \cite{PhysRevE.58.7475,MouZhangInPerp}
\begin{equation}
\label{SIForceEQ}
F_j = \left( -\frac{\partial f^{\text{el}}}{\partial Q_{pq,k}} Q_{pq,j} + (f^{\text{el}} + f^{\text{LdG}}) \delta_{jk} \right) \nu_k.
\end{equation}

We have provided simulated director fields for representative homeotropic and planar anchoring cases in Fig.~\ref{fig:simulations} \textbf{a, b, c} in the main text. The force in each simulation grid is calculated using (\ref{SIForceEQ}). To analyze the force distribution, we transform all simulated forces into spherical coordinates, choosing the \(-x\) direction (coaxis of the particle and bubble) as the north-pole direction. The \(X\)-\(Z\) section through the centers of the particle and bubble is defined as the azimuth angle \(\phi = 0\).

To obtain representative force distributions, we average over the azimuthal angle \(\phi\) and plot the distribution on the $xz$ plane on which the polar angle dependence of the pressure is given.
%rotate all forces with different \(\phi\) to the section \(\phi = 0\), and calculate the mean force for each polar angle \(\theta\). 
This gives the force distribution as a function of \(\theta\). The schematics in Fig.~\ref{fig:simulations} \textbf{d, e, f} are plotted on the basis of these results. We calculate the total force in the \(x\)-direction, \(F_{x}\) and report the value in Fig.~\ref{fig:simulations} in the main text.

In addition, we perform simulations to validate the expression for \(F_x^\text{defect} = -C K_1^p \sin({2\theta_b})\). Starting from a homeotropic anchoring dipole configuration, that is, a bulk hedgehog defect at the left pole of the particle, we observe a spontaneous transition to a Saturn ring defect at the equator (\(\theta_b = \pi/2\)). Taking six snapshots during this process, we calculate the force distribution (\Fref{fig:SI_particle}b-g and \(F_{x}\) on the particle. Fitting the data with $C = 0.5095$ and we have $K_1^p = 0.758$, we find an excellent agreement between the simulated \(F^\text{defect}_{x}\) and the analytical expression (see \Fref{fig:SI_particle}a).

\end{appendices}

\bibliography{cleaned_biblio} 

\providecommand{\noopsort}[1]{}\providecommand{\singleletter}[1]{#1}%

\begin{thebibliography}{32}
\ifx \bisbn   \undefined \def \bisbn  #1{ISBN #1}\fi
\ifx \binits  \undefined \def \binits#1{#1}\fi
\ifx \bauthor  \undefined \def \bauthor#1{#1}\fi
\ifx \batitle  \undefined \def \batitle#1{#1}\fi
\ifx \bjtitle  \undefined \def \bjtitle#1{#1}\fi
\ifx \bvolume  \undefined \def \bvolume#1{\textbf{#1}}\fi
\ifx \byear  \undefined \def \byear#1{#1}\fi
\ifx \bissue  \undefined \def \bissue#1{#1}\fi
\ifx \bfpage  \undefined \def \bfpage#1{#1}\fi
\ifx \blpage  \undefined \def \blpage #1{#1}\fi
\ifx \burl  \undefined \def \burl#1{\textsf{#1}}\fi
\ifx \doiurl  \undefined \def \doiurl#1{\url{https://doi.org/#1}}\fi
\ifx \betal  \undefined \def \betal{\textit{et al.}}\fi
\ifx \binstitute  \undefined \def \binstitute#1{#1}\fi
\ifx \binstitutionaled  \undefined \def \binstitutionaled#1{#1}\fi
\ifx \bctitle  \undefined \def \bctitle#1{#1}\fi
\ifx \beditor  \undefined \def \beditor#1{#1}\fi
\ifx \bpublisher  \undefined \def \bpublisher#1{#1}\fi
\ifx \bbtitle  \undefined \def \bbtitle#1{#1}\fi
\ifx \bedition  \undefined \def \bedition#1{#1}\fi
\ifx \bseriesno  \undefined \def \bseriesno#1{#1}\fi
\ifx \blocation  \undefined \def \blocation#1{#1}\fi
\ifx \bsertitle  \undefined \def \bsertitle#1{#1}\fi
\ifx \bsnm \undefined \def \bsnm#1{#1}\fi
\ifx \bsuffix \undefined \def \bsuffix#1{#1}\fi
\ifx \bparticle \undefined \def \bparticle#1{#1}\fi
\ifx \barticle \undefined \def \barticle#1{#1}\fi
\bibcommenthead
\ifx \bconfdate \undefined \def \bconfdate #1{#1}\fi
\ifx \botherref \undefined \def \botherref #1{#1}\fi
\ifx \url \undefined \def \url#1{\textsf{#1}}\fi
\ifx \bchapter \undefined \def \bchapter#1{#1}\fi
\ifx \bbook \undefined \def \bbook#1{#1}\fi
\ifx \bcomment \undefined \def \bcomment#1{#1}\fi
\ifx \oauthor \undefined \def \oauthor#1{#1}\fi
\ifx \citeauthoryear \undefined \def \citeauthoryear#1{#1}\fi
\ifx \endbibitem  \undefined \def \endbibitem {}\fi
\ifx \bconflocation  \undefined \def \bconflocation#1{#1}\fi
\ifx \arxivurl  \undefined \def \arxivurl#1{\textsf{#1}}\fi
\csname PreBibitemsHook\endcsname

%%% 1
\bibitem[\protect\citeauthoryear{Adler}{1975}]{adler1975chemotaxis}
\begin{barticle}
\bauthor{\bsnm{Adler}, \binits{J.}}:
\batitle{Chemotaxis in bacteria}.
\bjtitle{Annual review of biochemistry}
\bvolume{44}(\bissue{1}),
\bfpage{341}--\blpage{356}
(\byear{1975})
\end{barticle}
\endbibitem

%%% 2
\bibitem[\protect\citeauthoryear{Berg}{1975}]{berg1975chemotaxis}
\begin{barticle}
\bauthor{\bsnm{Berg}, \binits{H.C.}}:
\batitle{Chemotaxis in bacteria.}
\bjtitle{Annual review of biophysics and bioengineering}
\bvolume{4}(\bissue{00}),
\bfpage{119}--\blpage{136}
(\byear{1975})
\end{barticle}
\endbibitem

%%% 3
\bibitem[\protect\citeauthoryear{Hautier et~al.}{2009}]{hautier2009competition}
\begin{barticle}
\bauthor{\bsnm{Hautier}, \binits{Y.}},
\bauthor{\bsnm{Niklaus}, \binits{P.A.}},
\bauthor{\bsnm{Hector}, \binits{A.}}:
\batitle{Competition for light causes plant biodiversity loss after eutrophication}.
\bjtitle{Science}
\bvolume{324}(\bissue{5927}),
\bfpage{636}--\blpage{638}
(\byear{2009})
\end{barticle}
\endbibitem

%%% 4
\bibitem[\protect\citeauthoryear{Ballerini et~al.}{2008}]{ballerini2008interaction}
\begin{barticle}
\bauthor{\bsnm{Ballerini}, \binits{M.}},
\bauthor{\bsnm{Cabibbo}, \binits{N.}},
\bauthor{\bsnm{Candelier}, \binits{R.}},
\bauthor{\bsnm{Cavagna}, \binits{A.}},
\bauthor{\bsnm{Cisbani}, \binits{E.}},
\bauthor{\bsnm{Giardina}, \binits{I.}},
\bauthor{\bsnm{Lecomte}, \binits{V.}},
\bauthor{\bsnm{Orlandi}, \binits{A.}},
\bauthor{\bsnm{Parisi}, \binits{G.}},
\bauthor{\bsnm{Procaccini}, \binits{A.}}, \betal:
\batitle{Interaction ruling animal collective behavior depends on topological rather than metric distance: Evidence from a field study}.
\bjtitle{Proceedings of the national academy of sciences}
\bvolume{105}(\bissue{4}),
\bfpage{1232}--\blpage{1237}
(\byear{2008})
\end{barticle}
\endbibitem

%%% 5
\bibitem[\protect\citeauthoryear{Miller and Bassler}{2001}]{miller2001quorum}
\begin{barticle}
\bauthor{\bsnm{Miller}, \binits{M.B.}},
\bauthor{\bsnm{Bassler}, \binits{B.L.}}:
\batitle{Quorum sensing in bacteria}.
\bjtitle{Annual Reviews in Microbiology}
\bvolume{55}(\bissue{1}),
\bfpage{165}--\blpage{199}
(\byear{2001})
\end{barticle}
\endbibitem

%%% 6
\bibitem[\protect\citeauthoryear{Cates and Tailleur}{2015}]{cates2015motility}
\begin{barticle}
\bauthor{\bsnm{Cates}, \binits{M.E.}},
\bauthor{\bsnm{Tailleur}, \binits{J.}}:
\batitle{Motility-induced phase separation}.
\bjtitle{Annu. Rev. Condens. Matter Phys.}
\bvolume{6}(\bissue{1}),
\bfpage{219}--\blpage{244}
(\byear{2015})
\end{barticle}
\endbibitem

%%% 7
\bibitem[\protect\citeauthoryear{Howse et~al.}{2007}]{howse2007self}
\begin{barticle}
\bauthor{\bsnm{Howse}, \binits{J.R.}},
\bauthor{\bsnm{Jones}, \binits{R.A.}},
\bauthor{\bsnm{Ryan}, \binits{A.J.}},
\bauthor{\bsnm{Gough}, \binits{T.}},
\bauthor{\bsnm{Vafabakhsh}, \binits{R.}},
\bauthor{\bsnm{Golestanian}, \binits{R.}}:
\batitle{Self-motile colloidal particles: from directed propulsion to random walk}.
\bjtitle{Physical review letters}
\bvolume{99}(\bissue{4}),
\bfpage{048102}
(\byear{2007})
\end{barticle}
\endbibitem

%%% 8
\bibitem[\protect\citeauthoryear{Theurkauff et~al.}{2012}]{theurkauff2012dynamic}
\begin{barticle}
\bauthor{\bsnm{Theurkauff}, \binits{I.}},
\bauthor{\bsnm{Cottin-Bizonne}, \binits{C.}},
\bauthor{\bsnm{Palacci}, \binits{J.}},
\bauthor{\bsnm{Ybert}, \binits{C.}},
\bauthor{\bsnm{Bocquet}, \binits{L.}}:
\batitle{Dynamic clustering in active colloidal suspensions with chemical signaling}.
\bjtitle{Physical review letters}
\bvolume{108}(\bissue{26}),
\bfpage{268303}
(\byear{2012})
\end{barticle}
\endbibitem

%%% 9
\bibitem[\protect\citeauthoryear{Palacci et~al.}{2013}]{palacci2013living}
\begin{barticle}
\bauthor{\bsnm{Palacci}, \binits{J.}},
\bauthor{\bsnm{Sacanna}, \binits{S.}},
\bauthor{\bsnm{Steinberg}, \binits{A.P.}},
\bauthor{\bsnm{Pine}, \binits{D.J.}},
\bauthor{\bsnm{Chaikin}, \binits{P.M.}}:
\batitle{Living crystals of light-activated colloidal surfers}.
\bjtitle{Science}
\bvolume{339}(\bissue{6122}),
\bfpage{936}--\blpage{940}
(\byear{2013})
\end{barticle}
\endbibitem

%%% 10
\bibitem[\protect\citeauthoryear{Bricard et~al.}{2013}]{bricard2013emergence}
\begin{barticle}
\bauthor{\bsnm{Bricard}, \binits{A.}},
\bauthor{\bsnm{Caussin}, \binits{J.-B.}},
\bauthor{\bsnm{Desreumaux}, \binits{N.}},
\bauthor{\bsnm{Dauchot}, \binits{O.}},
\bauthor{\bsnm{Bartolo}, \binits{D.}}:
\batitle{Emergence of macroscopic directed motion in populations of motile colloids}.
\bjtitle{Nature}
\bvolume{503}(\bissue{7474}),
\bfpage{95}--\blpage{98}
(\byear{2013})
\end{barticle}
\endbibitem

%%% 11
\bibitem[\protect\citeauthoryear{Z{\"o}ttl and Stark}{2016}]{zottl2016emergent}
\begin{barticle}
\bauthor{\bsnm{Z{\"o}ttl}, \binits{A.}},
\bauthor{\bsnm{Stark}, \binits{H.}}:
\batitle{Emergent behavior in active colloids}.
\bjtitle{Journal of Physics: Condensed Matter}
\bvolume{28}(\bissue{25}),
\bfpage{253001}
(\byear{2016})
\end{barticle}
\endbibitem

%%% 12
\bibitem[\protect\citeauthoryear{Zhang et~al.}{2017}]{zhang2017active}
\begin{barticle}
\bauthor{\bsnm{Zhang}, \binits{J.}},
\bauthor{\bsnm{Luijten}, \binits{E.}},
\bauthor{\bsnm{Grzybowski}, \binits{B.A.}},
\bauthor{\bsnm{Granick}, \binits{S.}}:
\batitle{Active colloids with collective mobility status and research opportunities}.
\bjtitle{Chemical Society Reviews}
\bvolume{46}(\bissue{18}),
\bfpage{5551}--\blpage{5569}
(\byear{2017})
\end{barticle}
\endbibitem

%%% 13
\bibitem[\protect\citeauthoryear{Lavergne et~al.}{2019}]{lavergne2019group}
\begin{barticle}
\bauthor{\bsnm{Lavergne}, \binits{F.A.}},
\bauthor{\bsnm{Wendehenne}, \binits{H.}},
\bauthor{\bsnm{B{\"a}uerle}, \binits{T.}},
\bauthor{\bsnm{Bechinger}, \binits{C.}}:
\batitle{Group formation and cohesion of active particles with visual perception--dependent motility}.
\bjtitle{Science}
\bvolume{364}(\bissue{6435}),
\bfpage{70}--\blpage{74}
(\byear{2019})
\end{barticle}
\endbibitem

%%% 14
\bibitem[\protect\citeauthoryear{Fernandez-Rodriguez et~al.}{2020}]{fernandez2020feedback}
\begin{barticle}
\bauthor{\bsnm{Fernandez-Rodriguez}, \binits{M.A.}},
\bauthor{\bsnm{Grillo}, \binits{F.}},
\bauthor{\bsnm{Alvarez}, \binits{L.}},
\bauthor{\bsnm{Rathlef}, \binits{M.}},
\bauthor{\bsnm{Buttinoni}, \binits{I.}},
\bauthor{\bsnm{Volpe}, \binits{G.}},
\bauthor{\bsnm{Isa}, \binits{L.}}:
\batitle{Feedback-controlled active brownian colloids with space-dependent rotational dynamics}.
\bjtitle{Nature communications}
\bvolume{11}(\bissue{1}),
\bfpage{4223}
(\byear{2020})
\end{barticle}
\endbibitem

%%% 15
\bibitem[\protect\citeauthoryear{Mui{\~n}os-Landin et~al.}{2021}]{muinos2021reinforcement}
\begin{barticle}
\bauthor{\bsnm{Mui{\~n}os-Landin}, \binits{S.}},
\bauthor{\bsnm{Fischer}, \binits{A.}},
\bauthor{\bsnm{Holubec}, \binits{V.}},
\bauthor{\bsnm{Cichos}, \binits{F.}}:
\batitle{Reinforcement learning with artificial microswimmers}.
\bjtitle{Science Robotics}
\bvolume{6}(\bissue{52}),
\bfpage{9285}
(\byear{2021})
\end{barticle}
\endbibitem

%%% 16
\bibitem[\protect\citeauthoryear{Ben~Zion et~al.}{2023}]{ben2023morphological}
\begin{barticle}
\bauthor{\bsnm{Ben~Zion}, \binits{M.Y.}},
\bauthor{\bsnm{Fersula}, \binits{J.}},
\bauthor{\bsnm{Bredeche}, \binits{N.}},
\bauthor{\bsnm{Dauchot}, \binits{O.}}:
\batitle{Morphological computation and decentralized learning in a swarm of sterically interacting robots}.
\bjtitle{Science Robotics}
\bvolume{8}(\bissue{75}),
\bfpage{6140}
(\byear{2023})
\end{barticle}
\endbibitem

%%% 17
\bibitem[\protect\citeauthoryear{Geyer et~al.}{2019}]{geyer2019freezing}
\begin{barticle}
\bauthor{\bsnm{Geyer}, \binits{D.}},
\bauthor{\bsnm{Martin}, \binits{D.}},
\bauthor{\bsnm{Tailleur}, \binits{J.}},
\bauthor{\bsnm{Bartolo}, \binits{D.}}:
\batitle{Freezing a flock: Motility-induced phase separation in polar active liquids}.
\bjtitle{Physical Review X}
\bvolume{9}(\bissue{3}),
\bfpage{031043}
(\byear{2019})
\end{barticle}
\endbibitem

%%% 18
\bibitem[\protect\citeauthoryear{Lefranc et~al.}{2025}]{lefranc2025quorum}
\begin{botherref}
\oauthor{\bsnm{Lefranc}, \binits{T.}},
\oauthor{\bsnm{Dinelli}, \binits{A.}},
\oauthor{\bsnm{Fern{\'a}ndez-Rico}, \binits{C.}},
\oauthor{\bsnm{Dullens}, \binits{R.}},
\oauthor{\bsnm{Tailleur}, \binits{J.}},
\oauthor{\bsnm{Bartolo}, \binits{D.}}:
Quorum sensing and absorbing phase transitions in colloidal active matter.
arXiv preprint arXiv:2502.13919
(2025)
\end{botherref}
\endbibitem

%%% 19
\bibitem[\protect\citeauthoryear{Gu and Abbott}{2000}]{gu2000observation}
\begin{barticle}
\bauthor{\bsnm{Gu}, \binits{Y.}},
\bauthor{\bsnm{Abbott}, \binits{N.L.}}:
\batitle{Observation of saturn-ring defects around solid microspheres in nematic liquid crystals}.
\bjtitle{Physical Review Letters}
\bvolume{85}(\bissue{22}),
\bfpage{4719}
(\byear{2000})
\end{barticle}
\endbibitem

%%% 20
\bibitem[\protect\citeauthoryear{Mermin}{1981}]{mermin1981boojums}
\begin{botherref}
\oauthor{\bsnm{Mermin}, \binits{N.D.}}:
Boojums all the way through.
Physics Today,
46
(1981)
\end{botherref}
\endbibitem

%%% 21
\bibitem[\protect\citeauthoryear{Tasinkevych et~al.}{2012}]{tasinkevych2012liquid}
\begin{barticle}
\bauthor{\bsnm{Tasinkevych}, \binits{M.}},
\bauthor{\bsnm{Silvestre}, \binits{N.}},
\bauthor{\bsnm{Da~Gama}, \binits{M.T.}}:
\batitle{Liquid crystal boojum-colloids}.
\bjtitle{New journal of physics}
\bvolume{14}(\bissue{7}),
\bfpage{073030}
(\byear{2012})
\end{barticle}
\endbibitem

%%% 22
\bibitem[\protect\citeauthoryear{Romanczuk et~al.}{2012}]{romanczuk2012active}
\begin{barticle}
\bauthor{\bsnm{Romanczuk}, \binits{P.}},
\bauthor{\bsnm{B{\"a}r}, \binits{M.}},
\bauthor{\bsnm{Ebeling}, \binits{W.}},
\bauthor{\bsnm{Lindner}, \binits{B.}},
\bauthor{\bsnm{Schimansky-Geier}, \binits{L.}}:
\batitle{Active brownian particles: From individual to collective stochastic dynamics}.
\bjtitle{The European Physical Journal Special Topics}
\bvolume{202},
\bfpage{1}--\blpage{162}
(\byear{2012})
\end{barticle}
\endbibitem

%%% 23
\bibitem[\protect\citeauthoryear{Solon et~al.}{2015}]{solon2015active}
\begin{barticle}
\bauthor{\bsnm{Solon}, \binits{A.P.}},
\bauthor{\bsnm{Cates}, \binits{M.E.}},
\bauthor{\bsnm{Tailleur}, \binits{J.}}:
\batitle{Active brownian particles and run-and-tumble particles: A comparative study}.
\bjtitle{The European Physical Journal Special Topics}
\bvolume{224}(\bissue{7}),
\bfpage{1231}--\blpage{1262}
(\byear{2015})
\end{barticle}
\endbibitem

%%% 24
\bibitem[\protect\citeauthoryear{Conradi et~al.}{2010}]{musevicOptExpress2010}
\begin{barticle}
\bauthor{\bsnm{Conradi}, \binits{M.}},
\bauthor{\bsnm{Zorko}, \binits{M.}},
\bauthor{\bsnm{Mu{\v{s}}evi{\v{c}}}, \binits{I.}}:
\batitle{Janus nematic colloids driven by light}.
\bjtitle{Optics Express}
\bvolume{18}(\bissue{2}),
\bfpage{500}--\blpage{506}
(\byear{2010})
\end{barticle}
\endbibitem

%%% 25
\bibitem[\protect\citeauthoryear{Cui et~al.}{2017}]{CUI2017452}
\begin{barticle}
\bauthor{\bsnm{Cui}, \binits{J.}},
\bauthor{\bsnm{Long}, \binits{D.}},
\bauthor{\bsnm{Shapturenka}, \binits{P.}},
\bauthor{\bsnm{Kretzschmar}, \binits{I.}},
\bauthor{\bsnm{Chen}, \binits{X.}},
\bauthor{\bsnm{Wang}, \binits{T.}}:
\batitle{Janus particle-based microprobes: Determination of object orientation}.
\bjtitle{Colloids and Surfaces A: Physicochemical and Engineering Aspects}
\bvolume{513},
\bfpage{452}--\blpage{462}
(\byear{2017})
\end{barticle}
\endbibitem

%%% 26
\bibitem[\protect\citeauthoryear{Anthony et~al.}{2006}]{Granick}
\begin{barticle}
\bauthor{\bsnm{Anthony}, \binits{S.M.}},
\bauthor{\bsnm{Hong}, \binits{L.}},
\bauthor{\bsnm{Kim}, \binits{M.}},
\bauthor{\bsnm{Granick}, \binits{S.}}:
\batitle{Single-particle colloid tracking in four dimensions}.
\bjtitle{Langmuir}
\bvolume{22},
\bfpage{9812}--\blpage{9815}
(\byear{2006})
\end{barticle}
\endbibitem

%%% 27
\bibitem[\protect\citeauthoryear{Schindelin et~al.}{2012}]{Fiji}
\begin{barticle}
\bauthor{\bsnm{Schindelin}, \binits{J.}},
\bauthor{\bsnm{Arganda-Carreras}, \binits{I.}},
\bauthor{\bsnm{Frise}, \binits{E.}},
\bauthor{\bsnm{Kaynig}, \binits{V.}},
\bauthor{\bsnm{Longair}, \binits{M.}},
\bauthor{\bsnm{Pietzsch}, \binits{T.}},
\bauthor{\bsnm{Preibisch}, \binits{S.}},
\bauthor{\bsnm{Rueden}, \binits{C.}},
\bauthor{\bsnm{Saalfeld}, \binits{S.}},
\bauthor{\bsnm{Schmid}, \binits{B.}},
\bauthor{\bsnm{Tinevez}, \binits{J.-Y.}},
\bauthor{\bsnm{White}, \binits{D.J.}},
\bauthor{\bsnm{Hartenstein}, \binits{V.}},
\bauthor{\bsnm{Eliceiri}, \binits{K.}},
\bauthor{\bsnm{Tomancak}, \binits{P.}},
\bauthor{\bsnm{Cardona}, \binits{A.}}:
\batitle{Fiji: an open-source platform for biological-image analysis}.
\bjtitle{Nature Methods}
\bvolume{9},
\bfpage{676}--\blpage{682}
(\byear{2012})
\end{barticle}
\endbibitem

%%% 28
\bibitem[\protect\citeauthoryear{Allan et~al.}{2010--2023}]{Trackpy}
\begin{botherref}
\oauthor{\bsnm{Allan}, \binits{D.B.}},
\oauthor{\bsnm{Caswell}, \binits{T.}},
\oauthor{\bsnm{Keim}, \binits{N.C.}},
\oauthor{\bsnm{Wel}, \binits{C.M.}},
\oauthor{\bsnm{Verweij}, \binits{R.W.}}:
Soft-matter/trackpy: V0.6.1 (v0.6.1). Zenodo.
\url{https://doi.org/10.5281/zenodo.7670439}
\end{botherref}
\endbibitem

%%% 29
\bibitem[\protect\citeauthoryear{Qian and Sheng}{1998}]{PhysRevE.58.7475}
\begin{barticle}
\bauthor{\bsnm{Qian}, \binits{T.}},
\bauthor{\bsnm{Sheng}, \binits{P.}}:
\batitle{Generalized hydrodynamic equations for nematic liquid crystals}.
\bjtitle{Phys. Rev. E}
\bvolume{58},
\bfpage{7475}--\blpage{7485}
(\byear{1998})
\doiurl{10.1103/PhysRevE.58.7475}
\end{barticle}
\endbibitem

%%% 30
\bibitem[\protect\citeauthoryear{Mou and Zhang}{}]{MouZhangInPerp}
\begin{botherref}
\oauthor{\bsnm{Mou}, \binits{Z.}},
\oauthor{\bsnm{Zhang}, \binits{R.}}:
Q-tensor based continuum mechanics of surfaces immersed in nematic liquid crystals.
In prep.
\end{botherref}
\endbibitem

%%% 31
\bibitem[\protect\citeauthoryear{Tinevez et~al.}{2017}]{trackmate}
\begin{barticle}
\bauthor{\bsnm{Tinevez}, \binits{J.-Y.}},
\bauthor{\bsnm{Perry}, \binits{N.}},
\bauthor{\bsnm{Schindelin}, \binits{J.}},
\bauthor{\bsnm{Hoopes}, \binits{G.M.}},
\bauthor{\bsnm{Reynolds}, \binits{G.D.}},
\bauthor{\bsnm{Laplantine}, \binits{E.}},
\bauthor{\bsnm{Bednarek}, \binits{S.Y.}},
\bauthor{\bsnm{Shorte}, \binits{S.L.}},
\bauthor{\bsnm{Eliceiri}, \binits{K.W.}}:
\batitle{Trackmate: An open and extensible platform for single-particle tracking}.
\bjtitle{Methods}
\bvolume{115},
\bfpage{80}--\blpage{90}
(\byear{2017})
\end{barticle}
\endbibitem

%%% 32
\bibitem[\protect\citeauthoryear{Ahlers et~al.}{1994}]{PhysRevE.49.545}
\begin{barticle}
\bauthor{\bsnm{Ahlers}, \binits{G.}},
\bauthor{\bsnm{Cannell}, \binits{D.S.}},
\bauthor{\bsnm{Berge}, \binits{L.I.}},
\bauthor{\bsnm{Sakurai}, \binits{S.}}:
\batitle{Thermal conductivity of the nematic liquid crystal 4-n-pentyl-4'-cyanobiphenyl}.
\bjtitle{Phys. Rev. E}
\bvolume{49},
\bfpage{545}--\blpage{553}
(\byear{1994})
\doiurl{10.1103/PhysRevE.49.545}
\end{barticle}
\endbibitem

\end{thebibliography}

\end{document}